\documentclass[12pt]{article}
\usepackage{amssymb}
\usepackage{amsmath}
\usepackage[unicode,bookmarks,bookmarksopen,bookmarksopenlevel=2,colorlinks,linkcolor=blue,citecolor=green]{hyperref}
\usepackage{color}

\newtheorem{theorem}{Theorem}
\newtheorem{proposition}{Proposition}
\newtheorem{lemma}{Lemma}
\newtheorem{definition}{Definition}

\begin{document}

\title{\textbf{Classical Mechanical Systems with one-and-a-half Degrees of
Freedom and Vlasov Kinetic Equation}}
\author{Maxim V. Pavlov$^{1}$, Sergey P. Tsarev$^{2}$ \\
$^{1}$Sector of Mathematical Physics,\\
Lebedev Physical Institute of Russian Academy of Sciences,\\
Moscow, Leninskij Prospekt, 53\\
$^{2}$Siberian Federal University, \\
Institute of Space and Information Technologies,\\
26 Kirenski str., ULK-311, \\
Krasnoyarsk, 660074 Russia}
\date{}
\maketitle

\begin{flushright}
\textit{to honor of our Teacher Sergey Petrovich Novikov}
\end{flushright}

\begin{abstract}
We consider non-stationary dynamical systems with one-and-a-half degrees of
freedom. We are interested in algorithmic construction of rich classes of
Hamilton's equations with the Hamiltonian $H=p^2/2+V(x,t)$ which are
Liouville integrable. For this purpose we use the method of hydrodynamic
reductions of the corresponding one-dimensional Vlasov kinetic equation.

Also we present several examples of such systems with first integrals with
non-polynomial dependency w.r.t. to momentum.

The constructed in this paper classes of potential functions {$V(x,t)$}
which give integrable systems with one-and-a-half degrees of freedom are
parameterized by arbitrary number of constants.
\end{abstract}

\tableofcontents

\bigskip

\textit{keywords}: Vlasov kinetic equations, classical mechanics,
hydrodynamic type system, hydrodynamic reductions, Benney equations, L\"{o}%
wner equation, conservation law, Generalized Hodograph Method, Liouville
integrability.

\bigskip



\section{Introduction}

\label{sec-intro}

It is well known that many interesting and important classical mechanical
systems%
\begin{equation*}
\dot{q}_{i}=\frac{\partial H}{\partial p_{i}},\text{ \ }\dot{p}_{i}=-\frac{%
\partial H}{\partial q_{i}},\ \ i=1,\ldots ,n,
\end{equation*}%
determined by Hamiltonians
\begin{equation*}
H(\mathbf{q},\mathbf{p},t)=\frac{1}{2}\mathbf{p}^{2}+V(\mathbf{q},t),\quad
\mathbf{q}=(q_{1},\ldots ,q_{n}),\quad \mathbf{p}=(p_{1},\ldots ,p_{n}),
\end{equation*}%
are Liouville integrable, i.e. possess $n$ functions $F_{i}(\mathbf{q},%
\mathbf{p,}t)$ such that%
\begin{equation*}
\frac{dF_{i}}{dt}=0,\qquad \{F_{i},F_{j}\}_{\mathbf{p},\mathbf{q}}=0,
\end{equation*}%
where we have used the canonical Poisson bracket%
\begin{equation*}
\{F,G\}_{\mathbf{p},\mathbf{q}}=\underset{i=1}{\overset{n}{\sum }}\left(
\frac{\partial F}{\partial p_{i}}\frac{\partial G}{\partial q_{i}}-\frac{%
\partial F}{\partial q_{i}}\frac{\partial G}{\partial p_{i}}\right) .
\end{equation*}%
This means in particular that for every $F=F_i$%
\begin{equation*}
\frac{dF}{dt}\equiv \frac{\partial F}{\partial t}+\frac{\partial F}{\partial
q_{i}}\dot{q}_{i}+\frac{\partial F}{\partial p_{i}}\dot{p}_{i}=\frac{%
\partial F}{\partial t}+\frac{\partial F}{\partial q_{i}}\frac{\partial H}{%
\partial p_{i}}-\frac{\partial F}{\partial p_{i}}\frac{\partial H}{\partial
q_{i}}=0,
\end{equation*}%
or in a more compact form (Liouville equation)
\begin{equation*}
F_{t}=\{F,H\}_{\mathbf{p},\mathbf{q}}.
\end{equation*}

In this paper we restrict our considerations to the one-dimensional
non-autonomous case only. Such systems are usually called systems with
one-and-a-half degrees of freedom (\cite{kozlov,deryabin1997,MBialy}).
Everywhere below we identify $q_{1}=x$, $p_{1}=p$.

\begin{definition}
We call Hamilton's equations
\begin{equation}
\dot{x}=\frac{\partial H}{\partial p} \qquad \dot{p}=-\frac{ \partial H}{%
\partial x}  \label{hamilton}
\end{equation}
with the Hamiltonian function
\begin{equation}
H=\frac{p^{2}}{2}+V(x,t)  \label{1_5-freedom}
\end{equation}
\textbf{solvable (in hydrodynamic sense)} if there exists an additional
function $F(x,t,p)$ satisfying the Vlasov (collisionless Boltzmann) kinetic
equation (\cite{zakh}, \cite{Gibbons})
\begin{equation}
F_{t}-\{F,H\}_{p,x}=F_{t}+pF_{x}-F_{p}V_{x}=0,  \label{vlas}
\end{equation}
and the potential energy $V(x,t)$\ coincides with the zeroth moment $%
A^{0}(x,t)$ of the asymptotic expansion of the function $F(x,t,p)$ for $%
p\rightarrow \infty $:
\begin{equation}
F(x,t,p)=p+\frac{A^{0}(x,t)}{p}+\frac{A^{1}(x,t)}{p^{2}}+\frac{A^{2}(x,t)}{%
p^{3}}+ \ldots, \quad p\rightarrow \infty , \quad A^{0}(x,t)=V(x,t).
\label{expand}
\end{equation}
\end{definition}

Since Hamiltonian systems with one-and-a-half degrees of freedom solvable in
hydrodynamic sense have one additional first integral, they are Liouville
integrable. Thus we describe in this paper a subclass of Liouville
integrable Hamilton's equations using the method of hydrodynamic reductions.
Note that usually (for example in classical mechanics) (\ref{vlas}) is
interpreted as a first-order linear PDE with the unknown function $F$ and
\emph{fixed} potential $V(x,t)$. In our definition we impose a strong ansatz
$V=A^0(x,t)$. In many physical applications of various versions of the
Vlasov equation the quantities $A^{k}(x,t)$ (called ``moments'') are usually
introduced as integrals
\begin{equation}
A^{k}(x,t)=\overset{\infty }{\underset{-\infty }{\int }}p^{k}\Phi
(F(x,t,p))dp, \quad k=0,1,\ldots ,  \label{moment}
\end{equation}
where $\Phi (F)$ is an appropriate rapidly decreasing at infinities $%
p\rightarrow \pm \infty $ function such that the integrals are finite.
Certainly we can obtain the coefficients $A^{k}(x,t)$ in (\ref{expand}) as
residues at infinity $A^{k}(x,t)=\frac{1}{2\pi i}\oint p^{k} F(x,t,p)\, dp$
or choose another deformation of the contour in this integral. See
Appendix~A for more detail on possible relations of the coefficients in the
asymptotic expansion (\ref{expand}) and the integrals (\ref{moment}).

\textbf{Remark.} Below (see, for instance, (\ref{bkk}), (\ref{pui})) we
consider some very important solutions $F(x,t,p)$ which formally speaking do
not have the asymptotic behavior (\ref{expand}). However, equation (\ref%
{vlas}) is obviously invariant w.r.t. any point transformations $%
F(x,t,p)\mapsto f(F(x,t,p))$ with arbitrary function $f$ of one variable,
while all moments (\ref{moment}) are preserved after the appropriate change
of $\Phi (F)$. In all the cases treated below one can easily find the
modified $f(F(x,t,p))$ with the required asymptotic expansion (\ref{expand}%
). If the function $F$ has this asymptotic behavior then the ansatz $%
V=A^0(x,t)$ can be shown to be in fact a weak restriction on Liouville
integrable potentials $V(x,t)$ (see Appendix~A).

All this means that (see (\ref{1_5-freedom}), (\ref{vlas}), (\ref{expand}))
the Vlasov kinetic equation
\begin{equation}
F_{t}+pF_{x}-F_{p}V_{x}=0  \label{bolc}
\end{equation}%
in our approach is not a linear differential equation in partial derivatives
of the first order w.r.t. $F$, but a \emph{nonlinear integro-differential
equation}: we look for two unknown functions $F(x,t,p)$ and $V(x,t)$
simultaneously \emph{with the condition} $V(x,t)=A^{0}(x,t)$ in (\ref{expand}%
), so the complete problem of description of systems with one-and-a-half
degrees of freedom solvable in hydrodynamic sense is an integro-differential
equation for the function $F(x,t,p)$. The corresponding function $F(x,t,p)$
is called distribution function in plasma physics.

We prove in Section~\ref{sec-MHRed} that the variety of potentials
$V(x,t)$ solvable in hydrodynamic sense (via hydrodynamic reductions
with $N$ hydrodynamic variables) and the corresponding $F(x,t,p)$ is
parameterized by $2N$ arbitrary functions of a single variable for
arbitrary $N\geq 1$, where $N$ is the number of equations in
(\ref{rim}). Our approach is based on methods established in
\cite{GT} and developed in
\cite{algebra}, \cite{maksbenney}. Our construction of solvable potentials $%
V(x,t)$ by the method of hydrodynamic reductions provides essentially unique
$F(x,t,p)$ related to such a potential.

Integrability of the Vlasov kinetic equation was considered in a variety of
publications (see chronologically: \cite{Benney}, \cite{KM}, \cite{zakh},
\cite{Gibbons}, \cite{kodama}, \cite{krich}, \cite{bogd}, \cite{GT}, \cite%
{GibRai}, \cite{maksbenney}). A powerful method of hydrodynamic reductions
for the one-dimensional Vlasov type kinetic equation (including the Vlasov
kinetic equation itself) was developed in \cite{GT}, \cite{maksgen}, \cite%
{OPS}, while corresponding integrable hydrodynamic chains were investigated
in \cite{maksham}, \cite{FerMarsh}, \cite{MaksZyk}. We will describe it in
detail below and apply this method to description of a vast class of systems
with one-and-a-half degrees of freedom solvable in hydrodynamic sense.
Briefly speaking one should impose the following ansatz: $F(x,t,p)=\lambda (%
\mathbf{u},p)$ where $\mathbf{u}=(u^{1}(x,t),\ldots ,u^{N}(x,t))$ are
auxiliary \textquotedblleft hydrodynamic\textquotedblright\ unknown
functions. The possible forms of the function $\lambda(\mathbf{u},p)$ are
specified in our case explicitly in Section~\ref{sec-SNRed} (note that in
\cite{maksgen}, \cite{OPS} the method of hydrodynamic reductions requires $%
\lambda $ to be an unknown function as well). In this approach we will
always write $\lambda $ instead of $F$ if we impose this hydrodynamic
reduction ansatz. Two most interesting reductions $F(x,t,p)=\lambda (\mathbf{%
u},p)$ found before 1989 are:

\textbf{1}. \emph{The Bogdanov--Konopelchenko--Krichever reduction}:
\begin{equation}
\lambda =\frac{p^{N+1}}{N+1}+u^{0}p^{N-1}+ \ldots +u^{N-1}+\underset{m=1}{%
\overset{M}{\sum }}\left( \epsilon _{m}\ln (p-v^{m})+\underset{k=1}{\overset{%
K_{m}}{\sum }}\frac{w^{m,k}}{(p-v^{m})^{k}}\right) +\underset{l=1}{\overset{L%
}{\sum }}\tilde{\epsilon}_{l}\ln (p-\tilde{v}^{l}).  \label{bkk}
\end{equation}%
Here $\epsilon _{m}$ and $\tilde{\epsilon}_{l}$ are arbitrary constants, $%
N,M,L,K_{s}=0,1,\ldots $ and $u^{n},v^{m},w^{i,k},\tilde{v}^{l}$ are
functions of $x,t$ to be found.

\textbf{2}. \emph{The Puiseux type reduction}:
\begin{equation}
\lambda =\underset{n=1}{\overset{N+1}{\prod }}(p-a^{n}(x,t))^{\epsilon
_{n}}, \text{ \ }\underset{n=1}{\overset{N+1}{\sum }}\epsilon
_{n}a^{n}(x,t)=0,  \label{pui}
\end{equation}
where $N$ is an integer and $\epsilon _{n}$ are arbitrary constants (but $%
\sum \epsilon _{n}\neq 0$, because the leading term of the expansion of $%
\lambda $ for $p\rightarrow\infty$ must be a function of $p$ only). Only $%
a^{s}(x,t)$ for $s=1,\ldots, N$ are independent hydrodynamic variables.

Substitution of these expressions into (\ref{bolc}) leads to
corresponding hydrodynamic type systems (see \cite{dn}) for the
hydrodynamic variables $u^\bullet(x,t)$, $v^\bullet(x,t)$,
$w^\bullet(x,t)$ or $a^\bullet(x,t)$ in the aforementioned
reductions. We will use for simplicity the notation $\mathbf{u}
=(u^{1}(x,t), \ldots ,u^{N}(x,t))$ for the set of all hydrodynamic
variables in a given reduction. General solutions of corresponding
hydrodynamic type systems are parameterized by $N$ arbitrary
functions of a single variable (see \cite{Tsar,tsar91}). In the
first example (\ref{bkk}) we have the following \emph{fixed}
dependence $V(\mathbf{u})=u^{0}$. In the second example (\ref{pui}),
$V(\mathbf{a})=\frac{1}{2}\epsilon _{km}a^{k}a^{m}$ (see
(\ref{vpui}) below). Now let us suppose that we fix the precise
dependence $\lambda(\mathbf{u},p)$ (like (\ref{bkk}) or (\ref{pui}))
so $V(\mathbf{u})=A^0(\mathbf{u})$ in (\ref{expand}) is fixed as
well. In such a case the resulting potential functions
$V(x,t)=V(\mathbf{u}(x,t))$ of systems (\ref{hamilton}),
(\ref{1_5-freedom}) solvable in hydrodynamic sense \emph{are not
fixed}, they are parameterized by $N$ arbitrary functions of a
single variable for any $N \geq 1$, i.e. solutions of the
corresponding hydrodynamic type system for the hydrodynamic
variables $u^i(x,t)$ in the aforementioned reductions.

In a particular case (when all rational and logarithmic parts are removed), (%
\ref{bkk}) reduces to the form
\begin{equation}
\lambda _{(1)}=\frac{p^{N+1}}{N+1}+u^{0}p^{N-1}+\ldots +u^{N-1},  \label{a}
\end{equation}%
which is also equivalent to a particular case of (\ref{pui}) when all
constants $\epsilon _{n}=1$ (the condition $\sum \epsilon _{n}a^{n}=0$ in (%
\ref{pui}) means that the term $p^{N}$ vanishes in (\ref{a})). In such a
case (\ref{pui}) assumes the form
\begin{equation*}
\lambda _{(2)}=\underset{n=1}{\overset{N+1}{\prod }}(p-a^{n}(x,t)),\quad
\underset{n=1}{\overset{N+1}{\sum }}a^{n}(x,t)=0.
\end{equation*}%
Vlasov kinetic equation is invariant under \textit{any} point transformation
$F(x,t,p)\rightarrow \Phi (F(x,t,p))$, so $(N+1)\lambda _{(1)}=\lambda
_{(2)} $ give in fact the same reduction, the sets of hydrodynamic variables
$u^i$, $a^i$ are related by the obvious point transformation given by
Vieta's formulas. This is nothing but the well-known dispersionless limit of
the Gelfand--Dikij reduction for the remarkable Kadomtsev--Petviashvili
hierarchy (\cite{krich}). This unexpected relationship between ansatz (\ref%
{a}) for the integrable reductions of the Vlasov kinetic equation, classical
mechanical systems with one-and-a-half degrees of freedom and integrable
hydrodynamic type systems was implicitly or explicitly observed in a number
of publications, in particular \cite{kozlov,deryabin1997,MBialy, algebra}.
Let us give the following citation from \cite{kozlov}\footnote{%
translated from the Russian original by the authors}: \textquotedblleft It
has long been remarked that all the known first integrals of classical
mechanical systems are \textbf{polynomial}\footnote{%
boldface emphasis by the present authors} w.r.t. velocities (or functions of
such polynomials). This observation has no complete explanation yet\footnote{%
1989}. For this reason the analytical and geometrical nature of polynomial
integrals is of big interest\textquotedblright .

In this paper we \textbf{constructively} build a rich family of such
solvable potentials $V(x,t)$ with polynomial first integrals (\ref{a}).

In our construction the potential function $V(x,t)$ is one of the components
of a solution of some $N$ component hydrodynamic type system. In most cases
solutions of such systems break down in finite time. That means that the
potential function $V(x,t)$ may become singular.
The results of the papers \cite{kozlov,MBialy} on some classes of
nonsingular periodic potentials $V(x,t)$ are based exactly on this property
of quasilinear systems. Nevertheless (see for instance \cite{Chesn}) a class
of (probably piecewise analytic) nonsingular solutions can exist globally
for appropriate initial data.

Our \textit{second} contribution consists in \textbf{algorithmic}
construction of potentials $V(x,t)$ solvable in hydrodynamic sense with
\textbf{\ non-polynomial} first integrals $F(x,t,p)$ (cf. for example (\ref%
{bkk}) and (\ref{pui})), which depend on the momentum $p$ in a nontrivial
way.

This paper is organized as follows. In Section~\ref{sec-MHRed} we
briefly describe the method of hydrodynamic reductions. In
Section~\ref{sec-SNRed}, we consider polynomial and simplest
nonpolynomial in $p$ solutions of Vlasov kinetic equation
(\ref{bolc}). In Section~\ref{sec-HRedI} we construct solutions of
the waterbag and Puiseux type reductions by the Generalized
Hodograph Method. In Section~\ref{sec-similarity} we briefly
describe two types of similarity solutions for the Puiseux type
reductions and establish a remarkable link to the classical theory
of finite-gap potentials \cite{NZMP}. Section~\ref{sec-MTMS}
presents an alternative method of integration which produced two
interesting families of solvable potentials together with explicit
formulas for the solution of the corresponding Hamilton's equations
(\ref{hamilton}) without the need to use Liouville's theorem on
integrability in quadratures. In the Conclusion we remark that the
hydrodynamic reduction technique is applicable to a wider class of
Hamiltonians $H(V(x,t),p)$. Appendices~A, B, C and D contain some
technical details important for the method of hydrodynamic
reductions.

\section{Method of Hydrodynamic Reductions}

\label{sec-MHRed}

Substitution of (\ref{expand}) into (\ref{vlas}) with the restriction $%
V(x,t)=A^{0}(x,t)$ leads to the remarkable Benney hydrodynamic chain (see
\cite{Benney})
\begin{equation}
A_{t}^{k}+A_{x}^{k+1}+kA^{k-1}A_{x}^{0}=0,\text{ \ }k=0,1, \ldots
\label{cepochka}
\end{equation}
Substitution of (\ref{moment}) into (\ref{cepochka}) implies Vlasov kinetic
equation (\ref{vlas}) again (see Appendix~A).

According to the approach established in \cite{GT} we suppose that all $%
A^{k}(x,t)$, $V(x,t)$ have the form $A^{k}(x,t)=A^{k}(r^{1}(x,t),\ldots
,r^{N}(x,t))$, $V(x,t)=A^{0}(x,t)=V(r^{1}(x,t),\ldots ,r^{N}(x,t))$, where $%
V(\mathbf{r} )$, $A^{k}(\mathbf{r})$ are some fixed (unknown) functions of
the \textquotedblleft hydrodynamic variables\textquotedblright\ $r^{i}$ and
these variables $r^{i}=r^{i}(x,t)$ are \emph{arbitrary} solutions of an $N$
component hydrodynamic type system in diagonalized form (so the
corresponding \textquotedblleft hydrodynamic field
variables\textquotedblright\ $r^{i}$ are Riemann invariants of this
quasilinear system)
\begin{equation}
r_{t}^{i}+\mu ^{i}(\mathbf{r})r_{x}^{i}=0  \label{rim}
\end{equation}
integrable by the Generalized Hodograph Method (see \cite{Tsar,tsar91}). In
this case the functions $\mu^{i}(\mathbf{r})$ and $V(\mathbf{r})\equiv A^{0}(%
\mathbf{r})$ (see (\ref{expand}) and (\ref{bolc})) satisfy the so-called
Gibbons--Tsarev system (\cite{GT}) (here $\partial _{i}\equiv \partial
/\partial r^{i}$ )
\begin{equation}
\partial _{i}\mu ^{k}=\frac{\partial _{i}V}{\mu ^{i}-\mu ^{k}},\text{ \ }
\partial _{ik}^{2}V=2\frac{\partial _{i}V\partial _{k}V}{(\mu ^{i}-\mu
^{k})^{2}},\text{ }i\neq k,  \label{gt}
\end{equation}
while the function $F(x,t,p)=\lambda (\mathbf{r},p)$ satisfies the {\
(generalized) L\"{o}wner equations (\cite{GT})
\begin{equation}
\partial _{i}\lambda =\frac{\partial _{i}V}{p-\mu ^{i}}\lambda _{p},
\label{levner}
\end{equation}
whose compatibility conditions $\partial _{k}(\partial _{i}\lambda
)=\partial _{i}(\partial _{k}\lambda )$ lead to the Gibbons--Tsarev system (%
\ref{gt}). The celebrated L\"{o}wner equation initially appeared in 1923 as
an ordinary nonlinear differential equation describing deformations of
extremal univalent conformal mappings and was used in the solution of the
famous Bieberbach Conjecture in 1984 (see an exposition of the history of
this Conjecture in \cite{FomKuz} and its relation to hydrodynamic reductions
of Benney moment equations in \cite{GT}). Equations (\ref{gt}) and (\ref%
{levner}) were recently applied to the equations of Laplacian Growth,
Dirichlet Boundary Problem and Hele-Shaw problem (see for instance \cite%
{zabrod}). }

From (\ref{rim})--(\ref{levner}) we can determine the functional dimension
of the variety of potentials $V(x,t)$ integrable via hydrodynamic reductions
with $N$ hydrodynamic parameters $r^i(x,t)$. Namely, this variety is
parameterized by $2N$ functions of a single variable. First, the solutions
of the compatible system of equations (\ref{gt}) is parameterized by $2N$
functions of a single variable: the values of $V(\mathbf{r})$ on the
coordinate axes $r^i$ and the values of each $\mu ^{i}(\mathbf{r})$ on the
corresponding coordinate axis $r^i$ (the Goursat data for the system (\ref%
{gt})). The solutions $r^i(x,t)$ of (\ref{rim}) with fixed $\mu ^{i}(\mathbf{%
r})$ are parameterized by $N$ functions of a single variable and the
solutions $\lambda(\mathbf{r},p)$ of (\ref{levner}) with fixed $\mu ^{i}(%
\mathbf{r})$, $V(\mathbf{r})$ are parameterized by one function of a single
variable. However one can see that (\ref{levner}) essentially has only one
solution, the others are arbitrary functions of it: $\lambda \mapsto
f(\lambda)$. Also Riemann invariants $r^i$ for a given diagonalizable
hydrodynamic type system are fixed up to the change $r^i \mapsto f^i(r^i)$
so the variety of integrable potentials $V(x,t)=V(\mathbf{r}(x,t))$ is
parameterized by $2N$ functions of a single variable only.

As we have mentioned above we will use the notation $F(x,t,p)$ for the
conservation law only in the \emph{nonreduced} case and the notation $%
\lambda (\mathbf{r},p)$ for the same function in the case when a
finite-component hydrodynamic reduction of the Vlasov kinetic equation is
considered. According to the symmetric modification of the above method (see
\cite{algebra}), we can consider hydrodynamic type systems (\ref{rim})
written in the special (non-diagonal) conservative form with special
hydrodynamic variables $\mathbf{a}=(a^{1}(x,t),\ldots ,a^{N}(x,t))$:
\begin{equation}
a_{t}^{k}+\left( \frac{(a^{k})^{2}}{2}+V(\mathbf{a})\right) _{x}=0.
\label{semi}
\end{equation}%
Indeed, dividing all elements in Vlasov kinetic equation (\ref{bolc}) by $%
-F_{p}$ we get:
\begin{equation}
-\frac{F_{t}}{F_{p}}-p\frac{F_{x}}{F_{p}}+V_{x}=0,  \label{impli}
\end{equation}%
so according to the theorem about differentiation of implicit functions, one
can conclude that this equation assumes the form
\begin{equation}
p_{t}+\left( \frac{p^{2}}{2}+V\right) _{x}=0.  \label{gcl}
\end{equation}%
Here and below $p(x,t,F)$ is the inversion of the function $F(x,t,p)$ w.r.t.
$p$ (a solution of the implicit equation $F=F(x,t,p)$). It is a generating
function of conservation laws for Benney hydrodynamic chain (\ref{cepochka})
with respect to the parameter $F$. Let's choose $N$ arbitrary values $\xi
_{k}$ of this parameter $F$ and denote the corresponding functions $%
p(x,t,\xi _{k})$ as $a^{k}(x,t)$. Then $N$ copies of (\ref{gcl}) for
distinct values $\xi _{k}$ yield the hydrodynamic type system (\ref{semi}).
\emph{In this paper we will suppose that this set $a^{k}$ of the new
hydrodynamic variables is independent}. Let us study this problem in more
detail. Substitution of the asymptotic series
\begin{equation}
p(x,t,F)=F-\frac{H_{0}(x,t)}{F}-\frac{H_{1}(x,t)}{F^{2}}-\frac{H_{2}(x,t)}{%
F^{3}}-\ldots ,\text{ \ }F\rightarrow \infty  \label{asim}
\end{equation}%
(the inverted asymptotic series (\ref{expand})) into (\ref{gcl}) yields
Benney hydrodynamic chain (\ref{cepochka}) written in the conservative form
\begin{equation}
\partial _{t}H_{k}+\partial _{x}\left( H_{k+1}-\frac{1}{2}\underset{m=0}{%
\overset{k-1}{\sum }}H_{m}H_{k-1-m}\right) =0,\quad k=0,1,\ldots ,
\label{conserv}
\end{equation}%
where all conservation law densities $H_{k}$ are polynomials w.r.t. $A^{k}$.
For instance $%
H_{0}=A^{0},H_{1}=A^{1},H_{2}=A^{2}+(A^{0})^{2},H_{3}=A^{3}+3A^{0}A^{1}$.
According to the approach presented in \cite{GT}, all $N$ component
hydrodynamic reductions of Benney hydrodynamic chain (\ref{conserv}) can be
written choosing $N$ physical variables $H_{k}$, $k=0,1,\ldots ,N-1$ as an
independent set of hydrodynamic reduction variables, while all the other $%
H_{N-1+k}$ must be functions of this basic set $H_{0},\ldots H_{N-1}$ such
that all equations in (\ref{conserv}) must be consequences of the first $N$
of them. In such a case, we can introduce $N$ formal equalities (see (\ref%
{asim}))
\begin{eqnarray*}
a^{k}(H_{0},\ldots ,H_{N-1})\equiv p(\xi _{k}) &=& \\
\xi _{k}-\frac{H_{0}}{\xi _{k}}-\frac{H_{1}}{\xi _{k}^{2}}-\ldots -\frac{%
H_{N-1}}{\xi _{k}^{N}} &-&\frac{H_{N}(H_{0},\ldots ,H_{N-1})}{\xi _{k}^{N+1}}%
-\frac{H_{N+1}(H_{0},\ldots ,H_{N-1})}{\xi _{k}^{N+2}}-\ldots .
\end{eqnarray*}%
In this paper we consider the generic case: we assume that the point
transformation $\mathbf{H}\rightarrow \mathbf{a}(\mathbf{H})$ is invertible,
so the Jacobian $\partial a^{k}/\partial H_{j}$ is nondegenerate.
Nevertheless degenerate cases are also interesting and will be studied
elsewhere.

Thus each hydrodynamic reduction (\ref{semi}) has the generating function $p(%
\mathbf{a},\lambda )$ of conservation laws (cf. (\ref{gcl}))
\begin{equation}
p_{t}+\left( \frac{p^{2}}{2}+V(\mathbf{a})\right) _{x}=0,  \label{genred}
\end{equation}%
producing the infinite series (\ref{conserv}) of conservation law densities $%
H_{k}(\mathbf{a})$, where (cf. (\ref{asim}))
\begin{equation}
p(\mathbf{a},\lambda )=\lambda -\frac{H_{0}(\mathbf{a})}{\lambda }-\frac{%
H_{1}(\mathbf{a})}{\lambda ^{2}}-\frac{H_{2}(\mathbf{a})}{\lambda ^{3}}%
-\ldots ,\quad \lambda \rightarrow \infty .  \label{kruskal}
\end{equation}%
The function $V(\mathbf{a})$ satisfies the Gibbons--Tsarev system (cf. (\ref%
{gt}) in Riemann invariants)
\begin{equation}
(a^{i}-a^{k})\partial^2_{ik}V=\partial _{k}V\partial _{i}\left(
\sum_{m}\partial _{m}V\right) -\partial _{i}V\partial _{k}\left(
\sum_{m}\partial _{m}V\right) ,\quad \partial _{i}\equiv \partial /\partial
a^{i},\text{ \ }i\neq k.  \label{gibtsar}
\end{equation}%
System (\ref{gibtsar}) can be easily derived at least by two approaches.
First, let us consider the zeroth equation of Benney hydrodynamic chain (\ref%
{conserv}), i.e. $(H_{0}(\mathbf{a}))_{t}+(H_{1}(\mathbf{a}))_{x}=0$ or $%
\sum \partial _{k}H_{0}a_{t}^{k}+\sum \partial _{k}H_{1}a_{x}^{k}=0$.
Substituting $a_{t}^{k}$ from (\ref{semi}), and taking into account that
each factor of $a_{x}^{k}$ must vanish independently due to the assumption
that $a^{k}(x,t)$ are arbitrary solutions of (\ref{semi}), we conclude that $%
\partial _{k}H_{1}=\left( a^{k}+\sum \partial _{m}H_{0}\right) \partial
_{k}H_{0}$. Compatibility conditions $\partial _{i}(\partial
_{k}H_{1})=\partial _{k}(\partial _{i}H_{1})$ imply (\ref{gibtsar}), where
(as everywhere in this paper) $V(\mathbf{a})\equiv H_{0}(\mathbf{a})=A^{0}(%
\mathbf{a})$. Also, (\ref{genred}) yields $\sum \partial _{k}p(\mathbf{a}%
,\lambda )a_{t}^{k}+p(\mathbf{a},\lambda )\sum \partial _{k}p(\mathbf{a}%
,\lambda )a_{x}^{k}+\sum \partial _{k}V(\mathbf{a})a_{x}^{k}=0$. Repetition
of the above arguments leads to the L\"{o}wner equations (cf. (\ref{levner})
in Riemann invariants)
\begin{equation}
\partial _{i}p=\frac{\partial _{i}V}{p-a^{i}}\left( \sum_{m}\frac{\partial
_{m}V}{p-a^{m}}-1\right) ^{-1},  \label{law}
\end{equation}%
whose compatibility conditions $\partial _{k}(\partial _{i}p)=\partial
_{i}(\partial _{k}p)$ yield again Gibbons--Tsarev system (\ref{gibtsar}).

\textbf{Remark}. One can easily obtain L\"{o}wner equations for the function
$\lambda (\mathbf{a},p)$:
\begin{equation}
\partial _{i}\lambda =\frac{\partial _{i}V}{a^{i}-p}\left( \sum_{m}\frac{%
\partial _{m}V}{p-a^{m}}-1\right) ^{-1}\partial _{p}\lambda .
\label{loewner}
\end{equation}%
Details of the proof are given in Appendix~C. Dividing this equation by $%
\partial _{p}\lambda $ (cf. (\ref{impli})) and using the theorem about
differentiation of implicit functions, one arrives again to the L\"{o}wner
equations written in the form (\ref{law}). Nevertheless the L\"{o}wner
equations written in the form (\ref{loewner}) are more suitable for
integration. Indeed, introduce the auxiliary function $\varphi (\mathbf{a}%
,p) $ such that
\begin{equation}
\partial _{p}\lambda =\left( \sum_{m}\frac{\partial _{m}V}{p-a^{m}}-1\right)
\varphi (\mathbf{a},p).  \label{lev}
\end{equation}%
Then L\"{o}wner equations (\ref{loewner}) reduce to
\begin{equation}
\partial _{i}\lambda =\frac{\partial _{i}V}{a^{i}-p}\varphi (\mathbf{a},p).
\label{ner}
\end{equation}%
Thus integration of the L\"{o}wner equations written in the form (\ref%
{loewner}) is equivalent to computation of the integration factor $\varphi (%
\mathbf{a},p)$ subject to compatibility conditions following from (\ref{lev}%
), (\ref{ner}). We will see in Section~\ref{sec-SNRed} that in many cases we
are able to integrate (\ref{loewner}) completely in such a way.

The second order quasilinear system (\ref{gibtsar}) has a general solution
parameterized by $N$ arbitrary functions of a single variable. Currently we
do not have a constructive procedure to find this complete solution. In this
paper we will construct a finite-parametric family of solutions depending on
$N$ arbitrary constants for any $N\geq 1$ where $N$ is the number of
equations in (\ref{semi}) (see also \cite{maksbenney} for solutions with a
larger number of constant parameters).

Summarizing the necessary steps of the method of hydrodynamic reductions in
application to the problem of \emph{constructive} classification of
mechanical systems with one-and-a-half degrees of freedom solvable in
hydrodynamic sense and in order to give formulae for their potentials $%
V(x,t) $ and conservation laws $F(x,t,p)$ we sketch the following algorithm:

\begin{enumerate}
\item Once any solution $V(\mathbf{a})$ of (\ref{gibtsar}) is given, then
the corresponding semi-Hamiltonian hydrodynamic type system (\ref{semi}) is
fixed.

\item Compute the corresponding solution $\lambda (\mathbf{a},p)$ of L\"{o}%
wner equations (\ref{loewner}) (this problem is usually reduced to
computation of an integration factor $\varphi(\mathbf{a},p)$ and usually
found explicitly).

\item The corresponding system (\ref{semi}) possesses a general solution $%
a^{i}(x,t)$ parameterized by $N$ arbitrary functions of a single variable.

\item Thus, taking any given solution $V(\mathbf{a})$ of (\ref{gibtsar}) and
a general solution $a^{i}(x,t)$ of (\ref{semi}), we obtain the potential
functions $V(\mathbf{a}(x,t))$ as well as the additional functions $%
F(x,t,p)=\lambda (\mathbf{a}(x,t),p)$ parameterized by $N$ arbitrary
functions of a single variable and herewith we obtain infinitely many
Liouville integrable Hamilton's equations (\ref{hamilton}).
\end{enumerate}

Steps 1 and 3 of this algorithm need a close and detailed consideration,
since we do not have \emph{constructive} methods to obtain general solutions
of the systems (\ref{gibtsar}) and (\ref{semi}). The following Sections are
devoted to a way around this problem which produces in the cases considered
in this paper (and in many other cases, cf. \cite{gibb}, \cite{maksbenney})
explicit 
families of solutions.

Namely, in Section~\ref{sec-SNRed} we describe a few possible simple
solutions of (\ref{gibtsar}) together with corresponding solutions of (\ref%
{loewner}). A method of construction of rich families of solutions of the
system (\ref{semi}) is presented in Section~\ref{sec-HRedI}.

\section{Polynomial and Simplest Nonpolynomial Reductions}

\label{sec-SNRed}

As we have stated above, at this moment any regular procedure for
construction of solutions for the Gibbons--Tsarev system does not
exist. Nevertheless, some multi-parametric solutions can be found
easily. We give below a few explicit examples of such solutions. In
order to simplify the formulas we will give them modulo the obvious
point symmetry $a^i \mapsto \lambda a^i + \mu$ ($\lambda, \mu \in
\mathbb{R}$) of (\ref{gibtsar}).

\textbf{I}. Substitution of the ansatz $V(\mathbf{a})=\sum f_{m}(a^{m})$,
where $f_{k}(a^{k})$ are unknown functions, into (\ref{gibtsar}) yields the
following cases:

\textbf{I.1}. a particular $N$ parametric family of solutions (the so called
\emph{waterbag reduction}, see for instance \cite{GT}, \cite{LeiYu})
\begin{equation}
V(\mathbf{a})=\overset{N}{\underset{m=1}{\sum }}\epsilon _{m}a^{m},
\label{bag}
\end{equation}%
where all $\epsilon _{m}$ are arbitrary constants. Then L\"{o}wner equations
(\ref{loewner}) have the following solution
\begin{equation}
\lambda (\mathbf{a},p)=p-\overset{N}{\underset{m=1}{\sum }}\epsilon _{m}\ln
(p-a^{m}).  \label{water}
\end{equation}

\textbf{I.2}. A general solution
\begin{equation*}
V(\mathbf{a})=\overset{N}{\underset{m=1}{\sum }}\epsilon _{m}e^{a^{m}},
\end{equation*}%
where all $\epsilon _{m}$ are arbitrary constants. Then L\"{o}wner equations
(\ref{loewner}) have the following solution%
\begin{equation*}
\lambda (\mathbf{a},p)=-e^{-p}-\overset{N}{\underset{m=1}{\sum }}\epsilon
_{m}\overset{a^{m}-p}{\int }\frac{e^{q}dq}{q}.
\end{equation*}

\textbf{II}. A broader ansatz $V(\mathbf{a})=f(\Delta )$ (where $\Delta
=\sum f_{m}(a^{m})$ and $f_{k}(a^{k})$ are unknown functions) for (\ref%
{gibtsar}) yields
\begin{equation*}
V(\mathbf{a})=\ln (\Delta +\xi ),\text{\ \ \ }f_{k}^{\prime
}(a^{k})=\epsilon _{k}\exp \left( -\frac{(a^{k})^{2}}{2}\right) ,
\end{equation*}%
where $\xi $ and $\epsilon _{k}$ are arbitrary constants. Then L\"{o}wner
equations (\ref{loewner}) have the following solution%
\begin{equation*}
\lambda (\mathbf{a},p)=\xi\int e^{p^{2}/2}dp-\overset{N}{\underset%
{m=1}{\sum }}\epsilon _{m}\overset{a^{m}-p}{\int }\exp \left( -\frac{%
q^{2}+2pq}{2}\right) \frac{dq}{q}.
\end{equation*}%

\textbf{III}. A quadratic homogeneous polynomial ansatz $V(\mathbf{a})=\frac{%
1}{2}\epsilon _{km}a^{k}a^{m}$ leads to
\begin{equation}
V(\mathbf{a})=\frac{-1}{2(1+\epsilon )}\left[ \underset{m=1}{\overset{N}{%
\sum }}\epsilon _{m}(a^{m})^{2}+\left( \underset{m=1}{\overset{N}{\sum }}%
\epsilon _{m}a^{m}\right) ^{2}\right] ,\text{ \ }\epsilon =\underset{n=1}{%
\overset{N}{\sum }}\epsilon _{n}.  \label{vpui}
\end{equation}%
Then L\"{o}wner equations (\ref{loewner}) have the following solution (cf. (%
\ref{pui}))
\begin{equation}
\lambda (\mathbf{a},p)=\left( p+\underset{m=1}{\overset{N}{\sum }}\epsilon
_{m}a^{m}\right) \underset{n=1}{\overset{N}{\prod }}(p-a^{n})^{\epsilon
_{n}},\text{ \ }\underset{n=1}{\overset{N}{\sum }}\epsilon _{n}\neq -1.
\label{puiseux}
\end{equation}%
This is the so called \emph{Puiseux type reduction} (see for instance \cite%
{GT}). If all $\epsilon _{n}=1$, this is nothing but the dispersionless
limit of the Gelfand--Dikij reduction of the Kadomtsev--Petviashvili
hierarchy (see \cite{krich}). For Hamiltonian systems with one-and-a-half
degrees of freedom this class of polynomial integrals was studied in \cite%
{kozlov,deryabin1997,MBialy} where some results on existence of global \emph{%
nonsingular} periodic potentials were given. In this paper our approach is
essentially \emph{local}. If all $\epsilon _{n}=\pm 1$, this is the so
called \emph{Zakharov type reduction} (see \cite{zakh}); if $\epsilon
_{1}=-M $, ($M \neq N$) and all other $\epsilon _{n}=1$, this is the so
called \emph{Kodama reduction} (see \cite{algebra}). These three cases are
dispersionless limits of Krichever--Orlov reduction \cite{krich-red,orlov}
of the Kadomtsev--Petviashvili hierarchy, which can be obtained from (\ref%
{bkk}) if we remove the logarithmic terms.

More complicated reductions can be found also in \cite{maksbenney} and in a
set of publications \cite{gibb}.

\section{Hydrodynamic Reductions. Integrability}

\label{sec-HRedI}

In this Section we consider some constructive methods for integration of
hydrodynamic reductions (\ref{semi}) of Benney hydrodynamic chain (\ref%
{cepochka}). We illustrate in Sections~\ref{subsec-wb}, \ref{subsec-pui}
this construction on two examples: the \emph{waterbag reduction} (\ref{bag}%
), (\ref{water}) and the \emph{Puiseux type reduction} (\ref{vpui}), (\ref%
{puiseux}).

According to the Generalized Hodograph Method (see detail in \cite{Tsar},
\cite{tsar91}), any semi-Hamiltonian hydrodynamic type system (\ref{rim}),
i.e. a system whose characteristic velocities satisfy the integrability (or
the semi-Hamiltonian) property
\begin{equation*}
\partial _{j}\frac{\partial _{k}\mu ^{i}}{\mu ^{k}-\mu ^{i}}=\partial _{k}%
\frac{\partial _{j}\mu ^{i}}{\mu ^{j}-\mu ^{i}},\quad \partial _{i}\equiv
\partial /\partial r^{i},\text{ \ }i\neq j\neq k,
\end{equation*}%
possesses infinitely many commuting flows%
\begin{equation}
r_{\tau }^{i}=w^{i}(\mathbf{r})r_{x}^{i},  \label{com}
\end{equation}%
whose characteristic velocities are solutions of the linear system (again $%
\partial _{i}\equiv \partial /\partial r^{i}$)
\begin{equation}
\partial _{k}w^{i}=\frac{\partial _{k}\mu ^{i}}{\mu ^{k}-\mu ^{i}}%
(w^{k}-w^{i}),\text{ \ }i\neq k.  \label{lina}
\end{equation}%
The general solution of this compatible system depends on $N$ arbitrary
functions of a single variable. Then a generic solution $r^{i}(x,t)$ of
hydrodynamic type system (\ref{rim}) in a neighborhood of a generic point is
given in an implicit form by the algebraic system for the unknowns $%
r^{i}(x,t)$:
\begin{equation}
x-\mu ^{i}(\mathbf{r})\cdot t=w^{i}(\mathbf{r}),  \label{alg}
\end{equation}%
where $w^{i}(\mathbf{r})$ is a general solution of the compatible linear
system (\ref{lina}).

\textbf{Remark.} In arbitrary hydrodynamic variables $u^{i}(\mathbf{r})$
algebraic system (\ref{alg}) takes the form (see \cite{tsar91})
\begin{equation}
x\delta _{k}^{i}-tv_{j}^{i}(\mathbf{u})=w_{j}^{i}(\mathbf{u}),  \label{matr}
\end{equation}%
where the hydrodynamic type system (\ref{rim}) has the form%
\begin{equation*}
u_{t}^{i}=\sum_j v_{j}^{i}(\mathbf{u})u_{x}^{j},\text{ \ }i,j=1, \ldots ,N,
\end{equation*}%
while commuting hydrodynamic type systems (\ref{com}) have the form
\begin{equation*}
u_{\tau }^{i}=\sum_j w_{j}^{i}(\mathbf{u})u_{x}^{j},\text{ \ }i,j=1, \ldots
,N.
\end{equation*}%
In order to construct solutions of (\ref{semi}) we first need to prove the
Egorov property of Benney hydrodynamic chain (\ref{cepochka}). This property
is very important and many physical systems of hydrodynamic type integrable
by the Generalized Hodograph Method possess this property (cf. \cite{tsar91}%
, \cite{MaksTsar}). We need the following result suitable for investigation
of semi-Hamiltonian systems (cf.~\cite{MaksTsar}):

\begin{lemma}
\label{lemma-Egor} \label{lemma1} Any hydrodynamic reduction of Benney
hydrodynamic chain (\ref{cepochka}) has a Egorov pair of conservation laws $%
\big(f(\mathbf{u}(x,t))\big)_{t}= \big(h(\mathbf{u}(x,t))\big)_{x}$, $\big(h(%
\mathbf{u}(x,t))\big)_{t}=\big(g(\mathbf{u}(x,t))\big)_{x}$.
\end{lemma}

\textbf{Proof}. Indeed, two first conservation laws of (\ref{conserv}) are%
\begin{equation*}
\partial _{t}H_{0}+\partial _{x}H_{1}=0,\text{ \ }\partial _{t}H_{1}+\left(
H_{2}-\frac{1}{2}(H_{0})^{2}\right) _{x}=0,
\end{equation*}%
Any hydrodynamic reduction $H_{k}=H_{k}(\mathbf{u})$ of (\ref{conserv}) also
has these conservation laws so we can take $f=H_{0}(\mathbf{u}),h=-H_{1}(%
\mathbf{u})$, $g=H_{2}(\mathbf{u})-(H_{0}(\mathbf{u}))^{2}/2$.\hfill $%
\square $

Using the technique of \cite{MaksTsar} one easily proves that for
arbitrarily chosen conservation law density $h(\mathbf{r})$ of the original
system (in our case (\ref{rim})) an appropriately chosen commuting flow must
have a Egorov pair such that $f_{\tau }=h_{x}$, where $f=H_{0}$ (see
Appendix~B for the proof). Algebraic system (\ref{alg}), (or (\ref{matr}) in
arbitrary variables) can be written in the form (here $\partial
_{i}=\partial /\partial r^{i}$)
\begin{equation}
x-t\frac{\partial _{i}H_{1}}{\partial _{i}H_{0}}=\frac{\partial _{i}h}{%
\partial _{i}H_{0}}.  \label{temp}
\end{equation}%
Indeed, hydrodynamic type system (\ref{rim}) has a conservation law $%
\partial _{t}H_{0}+\partial _{x}H_{1}=0$, while the commuting hydrodynamic
system has the conservation law $f_{\tau }=h_{x}$. This means that $\partial
_{i}H_{0}r_{t}^{i}+\partial _{i}H_{1}r_{x}^{i}=0$ and $\partial
_{i}H_{0}r_{\tau }^{i} = \partial _{i}h\cdot r_{x}^{i}$. Taking into account
(\ref{rim}), (\ref{com}) and (\ref{alg}), one obtains (\ref{temp}).
Multiplying (\ref{temp}) by $\partial _{i}H_{0}dr^{i}$ and summing up, one
arrives at%
\begin{equation*}
xdH_{0}(\mathbf{r})-tdH_{1}(\mathbf{r})=dh(\mathbf{r}).
\end{equation*}%
Now we rewrite this equation after the invertible point transformation $(%
\mathbf{r})\rightarrow (\mathbf{a})$ as $xdH_{0}(\mathbf{a})-tdH_{1}(\mathbf{%
a})=dh(\mathbf{a})$, so the algebraic system (\ref{alg}) becomes%
\begin{equation}
x\frac{\partial H_{0}}{\partial a^{i}}-t\frac{\partial H_{1}}{\partial a^{i}}%
=\frac{\partial h}{\partial a^{i}}.  \label{hodo}
\end{equation}%
Taking into account that $\partial _{k}H_{1}=\left( a^{k}+\sum \partial
_{m}V\right) \partial _{k}V$ (see (\ref{cepochka}) and (\ref{semi}), here
again $\partial _{i}=\partial /\partial a^{i}$ and we remind that $V\equiv
H_{0}=A^{0}$) and substituting $p(\mathbf{a},\lambda )${\ instead of }$h(%
\mathbf{a})$, we obtain the algebraic system (see (\ref{law}))
\begin{equation*}
x\partial _{i}V-t\left( a^{i}+\sum \partial _{m}V\right) \partial _{i}V=%
\frac{\partial _{i}V}{p-a^{i}}\left( \sum \frac{\partial _{m}V}{p-a^{m}}%
-1\right) ^{-1},
\end{equation*}%
which is nothing but the diagonal part of the matrix algebraic system (\ref%
{matr}). All off-diagonal equations are compatible with the diagonal part (%
\cite{tsar91}).

So we proved:

\begin{theorem}
\label{theor1} An arbitrary hydrodynamic reduction (\ref{semi}) of Benney
hydrodynamic chain (\ref{cepochka}) has infinitely many particular solutions
$a^{i}(x,t)$ in the implicit form (here $\partial _{i}\equiv \partial
/\partial a^{i}$)
\begin{equation}
x-t\left( a^{i}+\underset{m=1}{\overset{N}{\sum }}\partial_{m}V\right) =
\frac{1}{p-a^{i}}\left( \underset{m=1}{\overset{N}{\sum }}\frac{\partial_{m}V%
}{p-a^{m}}-1\right) ^{-1},  \label{alga}
\end{equation}
where $p(\mathbf{a},\lambda )$ is the generating function of conservation
law densities (see (\ref{genred})).
\end{theorem}

Thus, once the function $V(\mathbf{a})$ is fixed (any solution of
Gibbons--Tsarev system (\ref{gibtsar})), the function $p( \mathbf{a},\lambda
)$ also is found as an inverse function to $\lambda (\mathbf{a},p)$ solving (%
\ref{loewner}) or computing the integrating factor $\varphi(\mathbf{a},p)$
in (\ref{lev}), (\ref{ner}). For the particular cases of $V(\mathbf{a})$
considered in Section~\ref{sec-SNRed} respective $\lambda (\mathbf{a},p)$
are explicitly given. Then the algebraic system (\ref{alga}) determines one
parametric family of solutions $a^{i}(x,t,\lambda )$ in implicit form and
simultaneously $V(x,t,\lambda)=V(\mathbf{a}(x,t,\lambda) )$. Thus we found
one parametric family of Hamilton's equations (\ref{hamilton}), which are
Liouville integrable.

In fact, using the Generalized Hodograph Method and the \emph{nonlinear
superposition principle} implied by this method (see below) we easily obtain
multiparametric families of solvable potentials. Namely expanding the
generating function $p(\mathbf{a},\lambda )$ at different points on the
Riemannian surface $p=p(\mathbf{a},\lambda )$ with the parameters $%
(p,\lambda)$ (for example when $p\rightarrow \infty$ or $p\rightarrow a^{i}$%
), one can construct infinite multiparametric series of new solutions $V(%
\mathbf{a}(x,t))$. Let us demonstrate this idea in detail.

\textbf{1}. \textit{Kruskal series}. Substitution of asymptotic expansion (%
\ref{kruskal}) into (\ref{genred}) leads to the Kruskal series of particular
conservation law densities $p_{0}^{k}(\mathbf{a})=H_{k}(\mathbf{a})$. They
can be found in quadratures. Indeed, we have an infinite series of
conservation laws (\ref{conserv}), where (let us remind) $H_{0}(\mathbf{a}%
)=A^{0}(\mathbf{a})=V(\mathbf{a})$:%
\begin{equation*}
(H_{0}(\mathbf{a}))_{t}+\left( H_{1}(\mathbf{a})\right) _{x}=0,\text{ \ }%
(H_{1}(\mathbf{a}))_{t}+\left( H_{2}(\mathbf{a})-\frac{1}{2}H_{0}^{2}(%
\mathbf{a})\right) _{x}=0,
\end{equation*}
\begin{equation*}
(H_{2}(\mathbf{a}))_{t}+\left( H_{3}(\mathbf{a})-H_{0}(\mathbf{a})H_{1}(%
\mathbf{a})\right) _{x}=0, \ldots .
\end{equation*}
Taking into account (\ref{semi}), we obtain%
\begin{equation}\label{Krusk-gen}
\partial _{k}H_{1}(\mathbf{a})=(a^{k}+\delta V)\partial _{k}V,\text{ \ \ }%
\partial _{k}H_{2}(\mathbf{a})=[(a^{k})^{2}+a^{k}\delta V+\sum a^{m}\partial
_{m}V+(\delta V)^{2}+V]\partial _{k}V,...
\end{equation}%
where $\delta =\sum \partial /\partial a^{m}$. Thus, once the potential
function $V(\mathbf{a})$ is given, all other higher Kruskal conservation law
densities are found by quadratures.

We call this asymptotic expansion 
Kruskal, because M.~Kruskal was first who introduced a similar expansion ($%
\lambda \rightarrow \infty $) for the KdV equation.

From (\ref{Krusk-gen}) we easily obtain a family of solvable
potentials $V(x,t)$ which is written in a compact form using
(\ref{hodo}) with $h(\mathbf{a})=\sum_s c_s H_s(\mathbf{a})$:
\begin{equation}
x-t(a^{k}+\delta V)=c_0+ c_1(a^{k}+\delta V) +
c_2\big((a^{k})^{2}+a^{k}\delta V+\sum a^{m}\partial _{m}V+(\delta
V)^{2}+V\big) + \ldots . \label{simple}
\end{equation}%
Namely suppose we have any solution $V(\mathbf{a})$ of
Gibbons--Tsarev system (\ref{gibtsar}). Then
solving algebraic system (\ref{simple}) with a fixed expression $V(\mathbf{a}%
)$, we can find $a^{k}(x,t)$ and $V(x,t)=V(\mathbf{a}%
(x,t))$. These solvable potentials are parameterized by arbitrary
number of constants $c_s$.

\textbf{2}. \textit{$N$ principal series}. Instead of asymptotic series (\ref%
{kruskal}) we can introduce $N$ expansions of $p(\mathbf{a},\lambda )$ at
the vicinities of $\lambda=\xi _{k}$. This means that we consider $N$ series
of conservation law densities%
\begin{equation}
p^{(k)}(\mathbf{a},\tilde{\lambda}_{(k)})=a^{k}+p_{1}^{k}(\mathbf{a})\tilde{%
\lambda}_{(k)}+p_{2}^{k}(\mathbf{a})\tilde{\lambda}_{(k)}^{2}+p_{3}^{k}(%
\mathbf{a})\tilde{\lambda}^{3}_{(k)}+ \ldots ,\text{ \ }k=1, \ldots ,N,
\label{princ}
\end{equation}%
so $p_{m}^{k}(\mathbf{a})$ are conservation law densities of hydrodynamic
type system (\ref{semi}), and $\tilde{\lambda}_{(k)}(\lambda )$ is a
corresponding local parameter at vicinity of each point $\lambda=\xi _{k}$, $%
p=a^{k}$. Substitution of each of these series into (\ref{genred}) yields (%
\ref{semi}) at the first step, while all higher conservation law densities $%
p_{m}^{k}(\mathbf{a})$ can be found at next steps in quadratures as we prove
in Appendix~D. This approach requires only $V(\mathbf{a})$ to be known
explicitly. Another algorithm to find the quantities $p_{m}^{k}(\mathbf{a})$
explicitly will be described in Sections~\ref{subsec-wb} and \ref{subsec-pui}
and requires the solution $\lambda (\mathbf{a},p)$ of (\ref{loewner}). These
$N$ series of conservation law densities $p_{m}^{k}(\mathbf{a})$ are
independent while the Kruskal series is their linear combination. However,
in some cases, the Kruskal series has its own interest, because
corresponding solutions are symmetric under arbitrary permutation of indices
of hydrodynamic variables $a^{k}$.

Once we found all these conservation law densities $p_{m}^{k}(\mathbf{a})$
and the Kruskal series $p_{0}^{k}(\mathbf{a})$, we can construct infinitely
many particular solutions parameterized by arbitrary number of constants $%
\sigma _{k}^{m}$ in (\ref{hodo}):
\begin{equation}
x\frac{\partial H_{0}}{\partial a^{i}}-t\frac{\partial H_{1}}{\partial a^{i}}%
=\frac{\partial }{\partial a^{i}}\left( \underset{k=1}{\overset{N}{\sum }}%
\underset{m=0}{\overset{\infty }{\sum }}\sigma _{k}^{m}p_{m}^{k}(\mathbf{a}%
)\right) ,  \label{summa}
\end{equation}%
or
\begin{equation}
x\frac{\partial H_{0}}{\partial a^{i}}-t\frac{\partial H_{1}}{\partial a^{i}}%
=\frac{\partial }{\partial a^{i}}\oint \varphi (\lambda )p(\mathbf{a}%
,\lambda )d\lambda ,  \label{inta}
\end{equation}%
where $\varphi (\lambda )$ and the contour can be chosen in many special
forms. Formulae (\ref{summa}) and (\ref{inta}) present \emph{the nonlinear
superposition principle} implied by the Generalized Hodograph Method.

Thus, once the function $V(\mathbf{a})$ is fixed (any solution of
Gibbons--Tsarev system (\ref{gibtsar})) and $\lambda (\mathbf{a},p)$ is
found from (\ref{loewner}) or (\ref{lev}), (\ref{ner}), algebraic system (%
\ref{summa}) determines multi-parametric families of solutions $a^{i}(x,t)$
in implicit form. By this way we simultaneously found $V(\mathbf{a}(x,t))$
and $F(x,t,p)=\lambda(\mathbf{a}(x,t),p)$. If the r.h.s. of algebraic system
(\ref{inta}) contains $N$ arbitrary functions $\varphi _{k}(\lambda )$, and
the contour consists of $N$ appropriate piecewise smooth curves (see, for
instance, \cite{krich}), then $a^{i}(x,t)$ depend on $N$ arbitrary functions
of a single variable. Then the potential function $V(\mathbf{a}(x,t))$ also
depends on $N$ arbitrary functions of a single variable. However, we cannot
describe such general solutions explicitly. Below we study in detail some
particular cases given in Section~\ref{sec-SNRed} and find rich
multiparametric families of solvable potentials.

\subsection{Waterbag Reduction}

\label{subsec-wb}

Waterbag hydrodynamic reduction (see (\ref{semi}) and (\ref{bag}))%
\begin{equation*}
a_{t}^{k}+\left( \frac{(a^{k})^{2}}{2}+\overset{N}{\underset{m=1}{\sum }}%
\epsilon _{m}a^{m}\right) _{x}=0
\end{equation*}%
has the Kruskal series of conservation laws (\ref{conserv}), where Kruskal
conservation law densities $H_{k}(\mathbf{a})$ are nonhomogeneous
polynomials w.r.t. $a^{k}$ (in a generic case, i.e. if $\sum \epsilon
_{m}\neq 0$). These polynomial expressions can be found by substitution of
asymptotic series (\ref{kruskal}) into (cf. (\ref{water})) the following
equation:
\begin{equation*}
\lambda _{(\infty )}-\overset{N}{\underset{m=1}{\sum }}\epsilon _{m}\cdot
\ln \lambda _{(\infty )}=p-\overset{N}{\underset{m=1}{\sum }}\epsilon
_{m}\ln (p-a^{m}).
\end{equation*}%
Here we perform a point transformation for the function $\lambda $:
$\lambda =\lambda _{(\infty )}-\sum \epsilon _{m}\cdot \ln \lambda
_{(\infty )}$ in order to have the asymptotic series of the form
(\ref{kruskal}). The first few Kruskal conservation law densities
are
$$
H_{0}(\mathbf{a})=\overset{N}{\underset{m=1}{\sum }}\epsilon _{m}a^{m},\text{
\ \ }H_{1}(\mathbf{a})=\frac{1}{2}\overset{N}{\underset{m=1}{\sum }}\epsilon
_{m}(a^{m})^{2}+\epsilon H_{0}(\mathbf{a}),
$$
$$H_{2}(\mathbf{a})=\frac{%
1}{3}\overset{N}{\underset{m=1}{\sum }}\epsilon _{m}(a^{m})^{3}+\overset{N}{%
\underset{m=1}{\sum }}\epsilon _{m}a^{m}H_{0}(\mathbf{a})+\epsilon H_{1}(%
\mathbf{a}),
$$
where $\epsilon =\overset{N}{\underset{m=1}{\sum }}\epsilon _{m}$.

$N$ principal series of conservation law densities can be found from
\begin{equation*}
\tilde{\lambda}_{(i)}=(p-a^{i})e^{-p/\epsilon _{i}}\underset{m\neq i}{\prod }%
(p-a^{m})^{\epsilon _{m}/\epsilon _{i}}
\end{equation*}%
for each index $i$ separately. Below we explicitly describe this procedure.
First we choose the corresponding local parameter $\tilde{\lambda}_{(i)}
=e^{-\lambda /\epsilon _{i}}$, then asymptotic series (\ref{princ}) is
applicable. Once a local parameter $\tilde{\lambda}_{(i)}$ is chosen so that
$\tilde{\lambda}_{(i)} \sim (p-a^{i})$, all conservation law densities $%
p_{m}^{k}(\mathbf{a})$ (see (\ref{princ})) can be obtained using the
Lagrange--B\"{u}rmann series (see, for instance, \cite{Lavr}) at the
vicinity of each singular point:

\begin{proposition}[Lagrange--B\"{u}rmann formula, \protect\cite{Lavr}]
\label{th-bur-lag} The analytic function
\begin{equation*}
y=y_{1}\cdot (x-x_{0})+y_{2}\cdot (x-x_{0})^{2}+y_{3}\cdot (x-x_{0})^{3}+
\ldots
\end{equation*}%
can be inverted ($y(x)\rightarrow x(y)$) as the Lagrange--B\"{u}rmann series
\begin{equation*}
x=x_{0}+x_{1}y+x_{2}y^{2}+x_{3}y^{3}+ \ldots ,
\end{equation*}%
whose coefficients are%
\begin{equation}
x_{n}=\frac{1}{n!}\underset{x\rightarrow x_{0}}{\lim }\frac{d^{n-1}}{dx^{n-1}%
}\left( \frac{x-x_{0}}{y}\right) ^{n}\text{, \ \ \ \ \ }n=1,2, \ldots
\label{ryad}
\end{equation}
\end{proposition}

This means that the conservation law densities $p_{m}^{k}(\mathbf{a})$ of
the waterbag reduction can be obtained with the aid of Lagrange--B\"{u}rmann
series:
\begin{equation*}
p_{n}^{i}=\frac{1}{n!}\frac{d^{n-1}}{d(a^{i})^{n-1}}\left(
e^{na^{i}/\epsilon _{i}}\underset{k\neq i}{\prod }(a^{i}-a^{k})^{-n\epsilon
_{k}/\epsilon _{i}}\right) ,\ \text{\ }n=1,2,\ldots ,\text{ \ }i=1,\ldots ,N.
\end{equation*}%
For instance, the first conservation law densities are%
\begin{equation*}
p_{1}^{i}=e^{a^{i}/\epsilon _{i}}\underset{k\neq i}{\prod }%
(a^{i}-a^{k})^{-\epsilon _{k}/\epsilon _{i}},\text{\ \ \ }p_{2}^{i}=\frac{%
e^{2a^{i}/\epsilon _{i}}}{\epsilon _{i}}\left( 1-\underset{n\neq i}{\sum }%
\frac{\epsilon _{n}}{a^{i}-a^{n}}\right) \underset{k\neq i}{\prod }%
(a^{i}-a^{k})^{-2\epsilon _{k}/\epsilon _{i}},\ldots
\end{equation*}%
Thus, multiparametric solutions can be found from (\ref{summa}). For example
in the simplest case (here $\kappa _{m}$ are arbitrary constants)
\begin{equation}
x\frac{\partial H_{0}}{\partial a^{i}}-t\frac{\partial H_{1}}{\partial a^{i}}%
=\frac{\partial }{\partial a^{i}}\left( \overset{N}{\underset{m=1}{\sum }}%
\kappa _{m}p_{1}^{m}\right) ,  \label{perv}
\end{equation}%
the corresponding algebraic system assumes the form%
\begin{equation*}
x-t(a^{i}+\epsilon )=\frac{\kappa _{i}p_{1}^{i}}{\epsilon _{i}^{2}}\left( 1-%
\underset{m\neq i}{\sum }\frac{\epsilon _{m}}{a^{i}-a^{m}}\right) +\underset{%
m\neq i}{\sum }\frac{\frac{\kappa _{m}}{\epsilon _{m}}p_{1}^{m}}{a^{m}-a^{i}}%
,\quad \epsilon =\sum \epsilon _{m}.
\end{equation*}

\subsection{Puiseux Type Reduction}

\label{subsec-pui}

Puiseux type hydrodynamic reduction (see (\ref{semi}) and (\ref{vpui}), here
$\epsilon =\sum \epsilon _{m}\neq -1$)%
\begin{equation}
a_{t}^{k}+\left( \frac{(a^{k})^{2}}{2}-\frac{1}{2(1+\epsilon )}\left[
\underset{m=1}{\overset{N}{\sum }}\epsilon _{m}(a^{m})^{2}+\left( \underset{%
m=1}{\overset{N}{\sum }}\epsilon _{m}a^{m}\right) ^{2}\right] \right) _{x}=0
\label{seux}
\end{equation}%
has the Kruskal series of conservation laws (\ref{conserv}), where Kruskal
conservation law densities $H_{k}(\mathbf{a})$ are homogeneous polynomials.
These polynomial expressions can be found by substitution of asymptotic
series (\ref{kruskal}) into (see (\ref{puiseux}))%
\begin{equation*}
\lambda =\left( p+\underset{m=1}{\overset{N}{\sum }}\epsilon
_{m}a^{m}\right) ^{\frac{1}{1+\epsilon }}\underset{n=1}{\overset{N}{\prod }}%
(p-a^{n})^{\frac{\epsilon _{n}}{1+\epsilon }}.
\end{equation*}%
Here we make a point transformation $\lambda \rightarrow \lambda
^{1+\epsilon }$, in order to obtain the asymptotic series (\ref{kruskal}).

$N$ principal series of conservation law densities are found from
\begin{equation*}
\tilde{\lambda}_{(i)}=(p-a^{i})\underset{m\neq i}{\prod }(p-a^{m})^{\epsilon
_{m}/\epsilon _{i}}\left( p+\underset{n=1}{\overset{N}{\sum }}\epsilon
_{n}a^{n}\right) ^{1/\epsilon _{i}}
\end{equation*}%
for each index $i$ separately. Here we choose the corresponding local
parameter $\tilde{\lambda}_{(i)}=\lambda ^{1/\epsilon _{i}}$, then
asymptotic series (\ref{princ}) is applicable. The conservation law
densities $p_{m}^{k}(\mathbf{a})$ of the Puiseux type reduction can be
obtained using (\ref{ryad})
\begin{equation*}
p_{n}^{i}=\frac{1}{n!}\frac{d^{n-1}}{d(a^{i})^{n-1}}\left( \underset{m\neq i}%
{\prod }(a^{i}-a^{m})^{-n\epsilon _{m}/\epsilon _{i}}\left( a^{i}+\underset{%
k=1}{\overset{N}{\sum }}\epsilon _{k}a^{k}\right) ^{-n/\epsilon _{i}}\right)
,\ \text{\ }n=1,2,\ldots ,\text{ \ }i=1,\ldots ,N.
\end{equation*}%
For instance, the first conservation law densities are%
\begin{equation}
p_{1}^{i}=\underset{m\neq i}{\prod }(a^{i}-a^{m})^{-\epsilon _{m}/\epsilon
_{i}}\left( a^{i}+\underset{k=1}{\overset{N}{\sum }}\epsilon
_{k}a^{k}\right) ^{-1/\epsilon _{i}},  \label{secon}
\end{equation}%
\begin{equation*}
p_{2}^{i}=-\frac{1}{\epsilon _{i}}\underset{m\neq i}{\prod }%
(a^{i}-a^{m})^{-2\epsilon _{m}/\epsilon _{i}}\left( a^{i}+\underset{k=1}{%
\overset{N}{\sum }}\epsilon _{k}a^{k}\right) ^{-2/\epsilon _{i}-1}\left[
1+\epsilon _{i}+\left( a^{i}+\underset{k=1}{\overset{N}{\sum }}\epsilon
_{k}a^{k}\right) \underset{n\neq i}{\sum }\frac{\epsilon _{n}}{a^{i}-a^{n}}%
\right] ,\ldots
\end{equation*}

\textbf{1}. \textit{Polynomial reduction (Dispersionless limit of the
Gelfand--Dikij reduction, see, for instance, \cite{krich})}. If all $%
\epsilon _{m}=1$, then Puiseux type reduction (\ref{puiseux}) becomes
polynomial
\begin{equation}
\lambda =\left( p+\underset{m=1}{\overset{N}{\sum }}a^{m}\right) \underset{%
n=1}{\overset{N}{\prod }}(p-a^{n}).  \label{poly}
\end{equation}%
As we mentioned before, this polynomial in $p$ case was studied in \cite%
{kozlov,deryabin1997,MBialy}. Unfortunately constructive results (formulae
for the potential $V(x,t)$ and the first integral $F(x,t,p)$) were obtained
in the context of classical mechanics only in a few cases. In this paper we
present a much wider class of such potentials and their first integrals. For
instance, $p_{1}^{i}$ in (\ref{secon}) for the Puiseux type reduction have
the homogeneity degree $K_{i}=1-\frac{1}{\epsilon _{i}}(\epsilon +1)$; in
the polynomial case all $K_{i}=-N$. The densities $p_{2}^{i}$ has the
homogeneity degree $K_{i}=1-\frac{2}{\epsilon _{i}}(\epsilon +1)$ for the
Puiseux type reduction; in the polynomial case, all $K_{i}=-2N-1$.
Corresponding expressions for conservation law densities are (see (\ref%
{secon}))
\begin{equation*}
p_{1}^{i}=\underset{m\neq i}{\prod }(a^{i}-a^{m})^{-1}\left( a^{i}+\underset{%
k=1}{\overset{N}{\sum }}
a^{k}\right) ^{-1},
\end{equation*}%
\begin{equation*}
p_{2}^{i}=-\underset{m\neq i}{\prod }(a^{i}-a^{m})^{-2}\left( a^{i}+\underset%
{k=1}{\overset{N}{\sum }}
a^{k}\right) ^{-3}\left[ 2+\left(
a^{i}+\underset{k=1}{\overset{N}{\sum }}
a^{k}\right) \underset{%
n\neq i}{\sum }\frac{1}{a^{i}-a^{n}}\right] ,\ldots
\end{equation*}

\textbf{2}. \textit{Zakharov type reduction (see, for instance, \cite{krich}
and \cite{zakh})}. If all $\epsilon _{m}=\pm 1$, then Puiseux type reduction
(\ref{puiseux}) assumes rational form with simple poles only, i.e.
\begin{equation*}
\lambda =\left( p+\sum_{m=1}^{N_{1}}a^{m}-\underset{n=1}{%
\overset{N_{2}}{\sum }}b^{n}\right) \frac{\underset{k=1}{\overset{N_{1}}{%
\prod }}(p-a^{k})}{\underset{s=1}{\overset{N_{2}}{\prod }}(p-b^{s})},
\end{equation*}%
where we introduced $N_{1}$ hydrodynamic variables $a^{k}(x,t)$ for $%
\epsilon _{k}=1$ and $N_{2}$ hydrodynamic variables $b^{m}(x,t)$ for $%
\epsilon _{m}=-1$. The Krichever reduction contains multiple poles.

\textbf{3}. \textit{Kodama reduction (see, for instance, \cite{krich} and
\cite{algebra})}. If all $\epsilon _{m}=1$ except $\epsilon _{1}=-M$ and $%
M\neq N$, then Puiseux type reduction (\ref{puiseux}) becomes rational with
one multiple pole:
\begin{equation*}
\lambda =\left( p+\underset{m=2}{\overset{N}{\sum }}a^{m}-Ma^{1}\right)
\frac{\underset{n=2}{\overset{N}{\prod }}(p-a^{n})}{(p-a^{1})^{M}}.
\end{equation*}

\section{Similarity Solutions}

\label{sec-similarity}

In this Section we consider a special but a very important sub-class of
solutions for Puiseux type reductions (\ref{vpui}), (\ref{puiseux}), (\ref%
{seux}).

Expansion (\ref{expand}) is invariant under the scaling
$F\rightarrow cF$, $p\rightarrow cp$ and $A^{k}\rightarrow
c^{k+2}A^{k}$, where $c$ is
arbitrary constant. Thus, it is easy to see that Benney hydrodynamic chain (%
\ref{cepochka}) admits similarity reductions $A^{k}(x,t)=t^{-(\beta
+1)(k+2)}B_{k}(z)$, where $z=xt^{\beta }$. Substitution of this ansatz
directly into Benney hydrodynamic chain (\ref{cepochka}) yields a chain of
ordinary differential equations%
\begin{equation*}
B_{k+1}^{\prime }(z)+\beta zB_{k}^{\prime }(z)-(\beta
+1)(k+2)B_{k}(z)+kB_{k-1}(z)B_{0}^{\prime }(z)=0,\text{ \ }k=0,1,2,...
\end{equation*}%
Hamilton's equations (\ref{hamilton}) are equivalent to a single
ordinary differential equation of a second order $\ddot{x}=-V_{x}$,
which reduces to the form
\begin{equation*}
\frac{d^{2}z}{d\tau ^{2}}-(2\beta +1)\frac{dz}{d\tau }+\beta (\beta
+1)z + \tilde V^{\prime }(z)=0,
\end{equation*}%
where $t=e^{\tau}$, $V(x,t)=A^{0}(x,t)=t^{-2(\beta +1)}B_{0}(z)$ and
$\tilde V(z)=B_{0}(z)$. This autonomous equation is equivalent to
the first order ordinary differential equation
\begin{equation}
q\frac{dq}{dz}-(2\beta +1)q+\beta (\beta +1)z+\tilde
V^{\prime}(z)=0, \label{first}
\end{equation}%
where $q=dz/d\tau $.
Our aim in this paper is to describe all functions $V(x,t)$ such
that the corresponding Hamilton's equations (\ref{hamilton}) are
solvable in hydrodynamic sense. In the above example, we would like
to find such $\tilde V(z)$.

For Puiseux type reductions (\ref{vpui}), (\ref{puiseux}), (\ref{seux}),
similarity solutions $a^{i}(x,t)=t^{-\beta -1}b^{i}(z)$ are determined by $N$
component non-autonomous system of ODEs%
\begin{equation}
\beta zb_{z}^{k}-(\beta +1)b^{k}+\partial _{z}\left( \frac{(b^{k})^{2}}{2}-%
\frac{1}{2(1+\epsilon )}\left[ \underset{m=1}{\overset{N}{\sum }}\epsilon
_{m}(b^{m})^{2}+\left( \underset{m=1}{\overset{N}{\sum }}\epsilon
_{m}b^{m}\right) ^{2}\right] \right) =0.  \label{ode-bz}
\end{equation}%
If $\beta =-1/2$, equation (\ref{first}) is easily integrable by the
method of separation of variables. Simultaneously, system
(\ref{ode-bz}) also integrates and yields $N$ algebraic equations
for $b^{k}(z)$:
\begin{equation*}
(b^{k})^{2}-zb^{k}-\beta _{k}=\frac{1}{(1+\epsilon )}\left[ \underset{m=1}{%
\overset{N}{\sum }}\epsilon _{m}(b^{m})^{2}+\left( \underset{m=1}{\overset{N}%
{\sum }}\epsilon _{m}b^{m}\right) ^{2}\right] ,
\end{equation*}%
where $\beta _{k}$ are arbitrary constants. Introducing new potential
function $V(\mathbf{b})=t\cdot V(\mathbf{a})$ (cf. (\ref{vpui})) we find
\begin{equation}
b^{k}(z)=\frac{z}{2}\pm \sqrt{\frac{z^{2}}{4}+\beta _{k}-2V(\mathbf{b}(z))}%
\,.  \label{bz_pui}
\end{equation}%
In fact we may avoid introduction of the new function $V(\mathbf{b})$ and
similar functions of the variables $\mathbf{b}$ below which differ by a
power of $t$ from $V(\mathbf{a})$ and other original ones, if we will
understand $V(\mathbf{b})$ as the result of \emph{formal substitution} of
the variables $b^{i}$ instead of $a^{i}$ directly into (\ref{vpui}) etc.
\emph{We will follow this understanding everywhere below in this Section},
for example for $A^{k}(\mathbf{b})$, $p_{s}^{k}(\mathbf{b})$ in (\ref%
{algsys-p}). The potential $V(\mathbf{b}(z))$ may be directly found from the
algebraic equation
\begin{equation*}
V(\mathbf{b}(z))=-\frac{1}{2(1+\epsilon )}\left[ \underset{m=1}{\overset{N}{%
\sum }}\epsilon _{m}(b^{m})^{2}+\left( \underset{m=1}{\overset{N}{\sum }}%
\epsilon _{m}b^{m}\right) ^{2}\right]
\end{equation*}%
where we should substitute $b^{k}(z)$ given by (\ref{bz_pui}). Then
\begin{equation*}
a^{m}(x,t)=t^{-1/2}\left( \frac{xt^{\beta }}{2}\pm \sqrt{\frac{(xt^{\beta
})^{2}}{4}+\beta _{m}-2V(\mathbf{b}(xt^{\beta }))}\right)
\end{equation*}%
and the corresponding first integral is given by (\ref{puiseux}):
\begin{equation*}
F(x,t,p)=\lambda (\mathbf{a}(x,t),p)=\left( p+\underset{m=1}{\overset{N}{%
\sum }}\epsilon _{m}a^{m}(x,t)\right) \underset{n=1}{\overset{N}{\prod }}%
(p-a^{n}(x,t))^{\epsilon _{n}}.
\end{equation*}

In the general case (when $\beta \neq -1/2$), integration of the above
non-autonomous system (\ref{ode-bz}) is a difficult problem. However, for
special values of the similarity exponent $\beta $ similarity solutions are
determined according to the Generalized Hodograph Method by an appropriate
choice of the commuting flows in (\ref{summa}). For instance, the simplest
similarity solution is determined by the algebraic system (see (\ref{perv})
and (\ref{secon}))
\begin{equation}
z\frac{\partial A^{0}(\mathbf{b})}{\partial b^{i}}-\frac{\partial A^{1}(%
\mathbf{b})}{\partial b^{i}}=\frac{\partial p_{s}^{k}(\mathbf{b})}{\partial
b^{i}},\text{ \ }i=1,\ldots ,N  \label{algsys-p}
\end{equation}%
for any indices $k$, $s$. The similarity exponent $\beta $ is determined
explicitly by the indices $k$, $s$ and the constants $\epsilon _{m}$. For
example, for $s=1$, 
$\beta =-\big(\epsilon _{k}+\epsilon +1\big)\big/\big(2\epsilon
_{k}+\epsilon +1\big)$. 

In the polynomial case (\ref{poly}), instead of algebraic system (\ref%
{algsys-p}) \textit{without free parameters}, $N$ parametric similarity
solutions can be presented, because all conservation law densities $%
p_{s}^{k}(\mathbf{b})$ with the same $s$ have the same homogeneity, i.e.
\begin{equation*}
z\frac{\partial {A}^{0}(\mathbf{b})}{\partial b^{i}}-\frac{\partial
{A}^{1}(\mathbf{b})}{\partial b^{i}}=\frac{\partial }{\partial b^{i}}%
\big[\sum_{m}\kappa _{m}p_{s}^{m}(\mathbf{b})\big],\quad i=1,\ldots ,N,
\end{equation*}%
(let us remind that all $\epsilon _{k}=1$ in the polynomial case). For
example when $s=1$, $
\beta =-\frac{N+2}{N+3}
$ and
\begin{equation*}
p_{1}^{i}(\mathbf{a})=\underset{m\neq i}{\prod }(a^{i}-a^{m})^{-1}\left(
a^{i}+\underset{k=1}{\overset{N}{\sum }}\epsilon _{k}a^{k}\right) ^{-1}.
\end{equation*}%
Similar computations can be made for arbitrary linear combination $%
\sum_{m}\kappa _{m}p_{s}^{m}(\mathbf{a})$ for any $s$.

Let us now consider another similarity reduction
$A^{k}=x^{k+2}B_{k}(t)$ (which may be obtained from the previous
reduction by appropriate limiting procedure with $\beta \rightarrow
\infty$). Then Benney hydrodynamic chain (\ref{cepochka}) reduces to
the form
\begin{equation*}
B_{k}^{\prime }(t)+(k+3)B_{k+1}(t)+2kB_{k-1}(t)B_{0}(t)=0,
\end{equation*}%
while Hamilton's equations (\ref{hamilton}) are equivalent to a
single ordinary differential equation of a second order
$\ddot{x}=-V_{x}$, which reduces to the linear ordinary differential
equation $\ddot{x}+2\tilde V(t)x=0$, where $V(x,t)=x^{2}\tilde V(t)$
and $\tilde V(t)=B_{0}(t)$. Again, integrability of the
corresponding Riccati equation in quadratures is an open problem in
the general case. By this reason, we also are interested in finding
functions $V(t)$ such that the corresponding Hamilton's equations
(\ref{hamilton}) will be Liouville integrable.

We seek similarity solutions of Puiseux type reductions in the form
$a^{i}(x,t)=b_{i}(t)x$. Substitution of this ansatz into
(\ref{seux}) yields $N$ component system of first order ordinary
differential equations
\begin{equation}
b_{k}^{\prime }(t)+b_{k}^{2}-\frac{1}{(1+\epsilon )}\left[ \underset{m=1}{%
\overset{N}{\sum }}\epsilon _{m}b_{m}^{2}+\left( \underset{n=1}{\overset{N}{%
\sum }}\epsilon _{n}b_{n}\right) ^{2}\right] =0.  \label{rikati}
\end{equation}%
After the potential substitution $b_{k}(t)=\psi _{k}^{\prime }(t)%
\big/\psi _{k}(t)$, system (\ref{rikati}) assumes the form
\begin{equation}\label{psi_k}
\psi _{k}^{\prime \prime }+2V(\mathbf{b})\psi _{k}=0,
\end{equation}%
where%
\begin{equation*}
V(\mathbf{b})=-\frac{1}{2(1+\epsilon )}\left[ \underset{m=1}{\overset{N}{%
\sum }}\epsilon _{m}(b^{m})^{2}+\left( \underset{m=1}{\overset{N}{\sum }}%
\epsilon _{m}b^{m}\right) ^{2}\right] =-\frac{1}{2(1+\epsilon )}\left[
\underset{m=1}{\overset{N}{\sum }}\epsilon _{m}\frac{\psi _{m}^{\prime ^{2}}%
}{\psi _{m}^{2}}+\left( \underset{m=1}{\overset{N}{\sum }}\epsilon _{m}\frac{%
\psi _{m}^{\prime }}{\psi _{m}}\right) ^{2}\right] .
\end{equation*}%
Since $\psi _{1}\psi _{k}^{\prime \prime }-\psi _{k}\psi
_{1}^{\prime \prime }=0$ for all $k>1$ we obtain
\begin{equation*}
\psi _{k}=g_{k}\psi _{1}\int \frac{dt}{\psi _{1}^{2}}+e_{k}\psi _{1},
\end{equation*}%
where $g_{k}$ and $e_{k}$ are arbitrary constants. Then ($k>1$)%
\begin{equation*}
b_{k}=\frac{\psi _{k}^{\prime }}{\psi _{k}}=\frac{\psi _{1}^{\prime }}{\psi
_{1}}+\frac{1}{\psi _{1}^{2}\int \frac{dt}{\psi _{1}^{2}}+s_{k}\psi _{1}^{2}}%
,
\end{equation*}%
where $s_{k}=e_{k}/g_{k}$.

Then (\ref{psi_k}) reduces to the form%
\begin{equation*}
-\frac{1}{2}pp^{\prime \prime }+\frac{1}{4}p^{\prime ^{2}}=\frac{1}{%
(1+\epsilon )}\left[ \frac{1}{4}\epsilon (1+\epsilon )p^{\prime
^{2}}-(1+\epsilon )pp^{\prime }\underset{n=2}{\overset{N}{\sum }}\frac{%
\epsilon _{n}}{r+s_{n}}+p^{2}\underset{m=2}{\overset{N}{\sum }}\frac{%
\epsilon _{m}}{(r+s_{m})^{2}}+p^{2}\left( \underset{n=2}{\overset{N}{\sum }}%
\frac{\epsilon _{n}}{r+s_{n}}\right) ^{2}\right] ,
\end{equation*}%
where $\psi _{1}(t)=p^{-1/2}(r)$ and the new independent variable
$r$ is defined by $r^{\prime }(t)=p(r)=(\psi_{1}(t))^{-2}$, so
\begin{equation*}
b_{1}=-\frac{1}{2}p^{\prime }(r),\text{ \ }b_{k}=-\frac{1}{2}p^{\prime }(r)+%
\frac{p}{r+s_{k}},\text{ \ }\epsilon =\underset{n=1}{\overset{N}{\sum }}%
\epsilon _{n}.
\end{equation*}%
Substitutions $p=z^{2(1+\epsilon )^{-1}}$ and%
\begin{equation*}
z(r)=y(r)\underset{m=2}{\overset{N}{\prod }}(r+s_{m})^{\epsilon
_{m}}
\end{equation*}%
lead to the differential equation $y^{\prime \prime }(r)=0$. Thus
(\ref{rikati}) is integrable in quadratures:
\begin{equation*}
t=t_{0}+\int (Ar+B)^{-2(1+\epsilon )^{-1}}\underset{m=2}{\overset{N}{\prod }%
}(r+s_{m})^{-2(1+\epsilon )^{-1}\epsilon _{m}}dr,
\end{equation*}%
and the expressions for $b_k(t)$ via $\psi _{1}(t)$ and $p(r)$ are
given here.

In the polynomial case (i.e. all $\epsilon _{m}=1$):%
\begin{equation*}
t=t_{0}+\int (Ar+B)^{-2(1+N)^{-1}}\underset{m=2}{\overset{N}{\prod }}%
(r+s_{m})^{-2(1+N)^{-1}}dr.
\end{equation*}%
If, for instance, $N=3$, then%
\begin{equation*}
t=t_{0}+\int \frac{dr}{\sqrt{(Ar+B)(r+s_{2})(r+s_{3})}}.
\end{equation*}%
So we see that (\ref{rikati}) is solvable in terms of the
Weierstrass $\wp $ function. This brings up a remarkable link
between the hydrodynamic reduction approach used here to obtain
some ``solvable'' potentials $\tilde V(t)$ and the classical theory
of finite-gap potentials \cite{NZMP}.

\section{Multi-Time Generalization. Explicit
Solutions}\label{sec-MTMS}


The approach presented in this paper can be extended to higher
number of \textquotedblleft time\textquotedblright\ variables.
According to \cite{Gibbons2} the Hamiltonian
$\tilde{H}=p^{3}/3+pV(x,y,t)+W(x,y,t)$ determines Hamilton's
equations with the new time variable $y$:
\begin{equation}
x_{y}=\frac{\partial \tilde{H}}{\partial p}=p^{2}+V(x,y,t),\text{ \ }p_{y}=-%
\frac{\partial \tilde{H}}{\partial x}=-p\frac{\partial V(x,y,t)}{\partial x}-%
\frac{\partial W(x,y,t)}{\partial x}.  \label{Ham-y}
\end{equation}%
This Hamiltonian system is compatible with the system
(\ref{hamilton}), (\ref{1_5-freedom}) (where the potential $V$ also
should be considered as a function of three variables $V(x,y,t)$) if
and only if $V(x,y,t)$, $W(x,y,t)$ satisfy (\ref{dkp_x}) given
below. Theory of such integrable pairs is of obvious interest. We
sketch below some aspects of this problem. If we will try to find an
appropriate definition of Liouville integrability one should
obviously start with the proper generalization of the definition of
conservation laws for such a pair of Hamiltonian systems with
``potentials'' $V(x,y,t)$, $W(x,y,t)$. Then the method of
hydrodynamic reductions may be used for construction of explicit
formulas for integrable potential pairs $V$, $W$. Corresponding
Vlasov type kinetic equation is
\begin{equation}
F_{y}+(p^{2}+V)F_{x}-F_{p}(pV_{x}+W_{x})=0.  \label{vlas_tk}
\end{equation}%
Again suppose now that $F(x,t,y,p)$ simultaneously satisfies
(\ref{vlas_tk}) and (\ref{bolc}). Their compatibility conditions
$(F_{t})_{y}=(F_{y})_{t}$ are equivalent to the following system of
equations for $V(x,y,t)$, $W(x,y,t)$:
\begin{equation}
V_{t}=-W_{x}, \quad \big(W_{t}-V_{y} - VV_{x}\big)_x=0,
\label{dkp_x}
\end{equation}%
This is a version of
the remarkable Lin--Reissner--Tsien equation (see \cite{LRT}) also
known as the Khokhlov--Zabolotskaya equation (see \cite{ZK}) or a
dispersionless limit of the Kadomtsev--Petviashvili equation (see
\cite{kp})
\begin{equation}
V_{t}=-W_{x},\text{ \ }W_{t}=V_{y}+VV_{x}.  \label{dkp}
\end{equation}%
Substitution of (\ref{expand}) into (\ref{vlas_tk}) yields the first
commuting flow from the Benney hierarchy (\cite{MaksEps}):
\begin{equation}
A_{y}^{k}+A_{x}^{k+2}+A^{0}A_{x}^{k}+(k+1)A^{k}A_{x}^{0}+kA^{k-1}A_{x}^{1}=0,
\text{ \ }k=0,1, \ldots ,  \label{cep-y}
\end{equation}%
where $V=A^{0}$ and $W=A^{1}$ (see also (\ref{moment})). Then all
the rest of further computations will be very similar to all that is
written here.
Functions $\lambda (\mathbf{a},p)$ are the same as well as all moments $%
A^{m}(\mathbf{a})$. However, a dependence with respect to
\textquotedblleft time\textquotedblright\ variable $y$ is given by
different hydrodynamic type
system (cf. (\ref{semi}))%
\begin{equation}
a_{y}^{k}+\left( \frac{(a^{k})^{3}}{3}+a^{k}V(\mathbf{a})+W(\mathbf{a}%
)\right) _{x}=0,  \label{cemi}
\end{equation}%
which commutes with (\ref{semi}) if and only if (\ref{dkp_x}) holds.
This means that moments $A^{m}(\mathbf{a})$ solve hydrodynamic
chains (\ref{cepochka}) and (\ref{cep-y}), where
functions $a^{i}(x,t,y)$ solve commuting hydrodynamic type systems (\ref%
{semi}) and (\ref{cemi}).

Thus, we just would like to mention here that the potential function
$V(x,t)$ of Hamilton's equations (\ref{hamilton}) can be interpreted
as a two-dimensional reduction of the function $V(x,y,t)=\partial
_{x}S(x,y,t)$, where we introduce $S(x,y,t)$ to simplify the
concept, since $S(x,y,t)$ is a solution of the Lin--Reissner--Tsien
equation written as a \emph{single}
three dimensional quasilinear equation of a second order%
\begin{equation*}
S_{tt}+S_{xy}+S_{x}S_{xx}=0
\end{equation*}%
and (see the first equation in (\ref{dkp})) $W(x,y,t)=-\partial _{t}S(x,y,t)$%
, while the first integral $F(x,y,t,p)$ satisfies two Vlasov type
kinetic
equations (see (\ref{vlas_tk}) and (\ref{bolc}))%
\begin{equation*}
F_{t}+pF_{x}-F_{p}S_{xx}=0,\text{ \ }%
F_{y}+(p^{2}+S_{x})F_{x}-F_{p}(pS_{xx}-S_{xt})=0.
\end{equation*}%
Let us remind that the both equations are nothing but $dF/dt=0$ and $dF/dy=0$%
, i.e. $F(x,y,t,p)=$const. A method of hydrodynamic reductions for
such three dimensional quasilinear equations of the second order was
developed in \cite{FerKar}.

Nevertheless, obviously, not all solutions of three dimensional
quasilinear
equation (see (\ref{dkp}))%
\begin{equation}\label{KhZ-V}
V_{tt}+(V_{y}+VV_{x})_{x}=0
\end{equation}%
can be obtained by the method of hydrodynamic reductions only.
However, if such a three dimensional quasilinear equation passes
this integrability test (i.e. possesses sufficiently many
hydrodynamic reductions), then this equation also can possess
particular solutions \textit{explicitly} parameterized by arbitrary
functions of a single variable. Moreover the two examples of
solvable potentials $V(x,y,t)$ given below have a remarkable
property: one can obtain \emph{explicit formulas for solutions of
the corresponding Hamilton's equations (\ref{hamilton}),
(\ref{1_5-freedom}) without the need to use Liouville's theorem on
integrability in quadratures}.

\textbf{1}. \textit{Manakov--Santini solution} (see \cite{MS}). A
particular class of solution $V(x,y,t)$ of (\ref{KhZ-V}) is given in
implicit form
\begin{equation}
x=2Vy-f(Vy^{1/2})-\frac{t^{2}}{4y},  \label{arbit}
\end{equation}%
where $f(U)$ is an arbitrary function. In order to solve the
Hamilton's equations (\ref{hamilton}), (\ref{1_5-freedom}),
(\ref{Ham-y}) for such potentials, we write them in the differential
form
\begin{equation}
dx=pdt+(p^{2}+V)dy,\text{ \ }dp=-V_{x}dt+(-pV_{x}+V_{t})dy
\label{razdif}
\end{equation}%
Under the substitutions $y=z^{2},t=z\tau ,zV=U$, (\ref{razdif}) read
\begin{equation}
dx=pzd\tau +(2zp^{2}+p\tau +2U)dz,\text{ \ }dp=-[2z-f^{\prime
}(U)]^{-1}d\tau -2[2z-f^{\prime }(U)]^{-1}pdz,  \label{dvadif}
\end{equation}%
while (\ref{arbit}) becomes
\begin{equation}
x=2zU-f(U)-\frac{\tau ^{2}}{4}.  \label{arbi}
\end{equation}%
Substitution (\ref{arbi}) into (\ref{dvadif}) yields%
\begin{equation}
dU=-\frac{q}{2}dp,\text{ \ }dq=f^{\prime }(U)dp,  \label{totaldif}
\end{equation}%
where $q=\tau +2pz$, and we consider now the function $U$ as the
function of two variables: $U(z,\tau ) = U(x(z,\tau),z,\tau )$
instead of $U(x,z,\tau)$ found from (\ref{arbi}). Thus, we obtain
\begin{equation}\label{pq-Man}
q=\pm 2\sqrt{E-f(U)},\text{ \ \ }p=p_{0}\mp \int
\frac{dU}{\sqrt{E-f(U)}},
\end{equation}%
where $p_{0}$ and $E$ are integration constants. Taking into account
(\ref{totaldif}) and $q=\tau +2pz$, we obtain $U(z,\tau ) =
U(x(z,\tau),z,\tau )$ from the equation
\begin{equation}
\pm 2\sqrt{E-f(U)}=\tau +2z\left( p_{0}\mp \int \frac{dU}{\sqrt{E-f(U)}}%
\right) .  \label{kvadrat}
\end{equation}%
So finally we obtain the solution of the two-time Hamilton's
equations
\begin{equation*}
x(z,\tau )=2zU(z,\tau )-f(U(z,\tau ))-\frac{\tau ^{2}}{4},
\end{equation*}%
where $U(z,\tau )$ is given by (\ref{kvadrat}); $p(z, \tau)$ is
found from (\ref{pq-Man}).

\textbf{2}. \textit{Simple wave solution}. The solution $V(x,y,t)$
of (\ref{KhZ-V}) is determined by
\begin{equation}
x=Q(V)-F(V)t+(V+F^{2}(V))y,  \label{simplewave}
\end{equation}%
where $Q(V)$ and $F(V)$ are arbitrary functions. Then (\ref{razdif})
read
\begin{equation*}
\lbrack Q^{\prime }(V)-F^{\prime }(V)t+(1+2F(V)F^{\prime
}(V))y]dV=(p+F(V))[dt+(p-F(V))dy],
\end{equation*}%
\begin{equation}\label{SimpleWave-d2}
-[Q^{\prime }(V)-F^{\prime }(V)t+(1+2F(V)F^{\prime
}(V))y]dp=dt+(p-F(V))dy.
\end{equation}%
These two differentials reduce to
\begin{equation*}
(p+F(V))dp+dV=0,
\end{equation*}%
\begin{equation*}
-[Q^{\prime }(V)-F^{\prime }(V)t+(1+2F(V)F^{\prime
}(V))y]dp=dt+(p-F(V))dy.
\end{equation*}%
The first of them means that $p=P(V)$, where $P(V)$ is a solution of
Abel equation
\begin{equation}\label{SimpleWave-Abel}
(P(V)+F(V))P^{\prime }(V)=-1.
\end{equation}
However, since $F(V)$ is an arbitrary
function, we can express $F(V)$ via $P(V)$, i.e.%
\begin{equation*}
F(V)=-\frac{1}{P^{\prime }(V)}-P(V).
\end{equation*}%
Then the differential (\ref{SimpleWave-d2}) can be integrated in
quadratures (with one effective integration constant, the second one
is hidden in (\ref{SimpleWave-Abel}))
\begin{equation*}
t+(2P(V)+\frac{1}{P^{\prime }(V)})y+P^{\prime }(V)e^{-\int P^{\prime
^{2}}(V)dV}\int Q^{\prime }(V)e^{\int P^{\prime ^{2}}(V)dV}dV=0.
\end{equation*}%
Thus we can find the solution $V(y,t)=V(x(y,t),y,t)$ of this
equation in implicit form. Then (\ref{simplewave}) yields the
solution $x(y,t)$ of (\ref{hamilton}), (\ref{1_5-freedom}),
(\ref{Ham-y}):
\begin{equation*}
x(y,t)=Q(V)+\left( \frac{1}{P^{\prime }(V)}+P(V)\right) t+\left[
V+\left( \frac{1}{P^{\prime }(V)}+P(V)\right) ^{2}\right] y.
\end{equation*}
The momentum $p(y,t)$ is found from the relation $p=P(V)$.

\section{Conclusion}

\label{sec-concl}

We have constructed a few multiparametric families of potentials $V(x,t)$
with integrals $F(x,t,p)$ which are either polynomial or non-polynomial in $%
p $. There is a strong evidence (cf. \cite{GT}) that such families are \emph{%
locally dense} in the functional space of all potentials. Unfortunately we
do not have a possibility to go into the necessary details here.

In this paper we considered Hamilton's equations (\ref{hamilton}),
determined by the \emph{classical} Hamiltonian function (\ref{1_5-freedom}).
They are equivalent to a single equation $\ddot{x}=-V_{x}$. Now we would
like to emphasize that our approach is applicable for Hamilton's equations (%
\ref{hamilton}) with Hamiltonian function $H(x,t,p)$ of much more general
form than (\ref{1_5-freedom}). This is based on the following results.

A complete classification of Vlasov type kinetic equations (cf. (\ref{vlas}%
))
\begin{equation*}
F_{t}-\{F,H\}=F_{t}+H_{p}F_{x}-F_{p}H_{x}=0
\end{equation*}%
integrable by the method of hydrodynamic reductions was presented in
\cite{OPS} for the Hamiltonian functions $H(V(x,t),p)$. First three
simplest
cases (see also \cite{maksgen}) have the form%
\begin{equation*}
H=Q_{1}(p)+V(x,t),\text{ \ }H=Q_{2}(p)+pV(x,t),\text{ \ }H=Q_{3}(p)V(x,t),
\end{equation*}%
where $Q_{i}(p)$ are arbitrary solutions of the equations
$Q_{1}^{\prime \prime }=\alpha Q_{1}^{\prime ^{2}}+\beta
Q_{1}^{\prime }+\gamma $, $pQ_{2}^{\prime \prime }=\alpha
Q_{2}^{\prime ^{2}}+\beta Q_{2}^{\prime }+\gamma $,
$Q_{3}Q_{3}^{\prime \prime }=\alpha Q_{3}^{\prime ^{2}}+\beta
Q_{3}^{\prime }+\gamma $ and $\alpha ,\beta ,\gamma $ are arbitrary
constants (if $Q_{1}(p)=p^{2}/2$, this is nothing but the case
considered in this paper). Corresponding analogues of L\"{o}wner
equation and Gibbons--Tsarev system were derived in \cite{OPS},
\cite{maksgen}. Following the approach presented here, one can
extract infinitely many particular solutions of L\"{o}wner equation
and Gibbons--Tsarev system and thus construct infinitely many
Hamilton's equations solvable in hydrodynamic sense.

\section*{Appendix A (Asymptotic Expansion and Moments)}

\addcontentsline{toc}{section}{Appendix A (Asymptotic Expansion and
Moments)}

As we mentioned in Introduction the distribution function $F(x,t,p)$ in our
approach satisfies the Vlasov (collisionless Boltzmann) kinetic equation (%
\cite{zakh}, \cite{Gibbons})
\begin{equation}
F_{t}+pF_{x}-F_{p}V_{x}=0,  \label{vlas-bis}
\end{equation}%
where the potential energy $V(x,t)$ coincides with the zeroth moment $%
A^{0}(x,t)$ of the asymptotic expansion of the function $F(x,t,p)$ for $%
p\rightarrow \infty $:%
\begin{equation}
F(x,t,p)=p+\frac{A^{0}(x,t)}{p}+\frac{A^{1}(x,t)}{p^{2}}+\frac{A^{2}(x,t)}{%
p^{3}}+\ldots ,\quad p\rightarrow \infty .  \label{expand-bis}
\end{equation}%
On the other hand in many mechanical and physical applications the moments $%
A^{k}$ are defined as
\begin{equation}
A^{k}(x,t)=\overset{\infty }{\underset{-\infty }{\int }}p^{k}\Phi
(F(x,t,p))dp,  \label{moment-bis}
\end{equation}%
where $\Phi (F)$ is an appropriate rapidly decreasing at infinities $%
p\rightarrow \pm \infty $ function such that the integrals are finite.

In this Appendix we study the relation of (\ref{vlas-bis}) with the
expansion (\ref{expand-bis}) on one hand and the same equation (\ref%
{vlas-bis}) associated with (\ref{moment-bis}) on the other hand.

First, we start with the pair (\ref{vlas-bis})+(\ref{expand-bis}). We will
consider an even more general asymptotic behavior at infinity $p\rightarrow
\infty $
\begin{equation}
F(x,t,p)=a_{-2}(x,t)p+a_{-1}(x,t)+\frac{a_{0}(x,t)}{p}+\frac{a_{1}(x,t)}{%
p^{2}}+\ldots .  \label{laurent}
\end{equation}%
Direct substitution of this expansion into (\ref{vlas-bis}) leads to an
infinite series of equations:
\begin{equation}
a_{-2,x}=0,  \label{ras}
\end{equation}%
\begin{equation}
a_{-2,t}+a_{-1,x}=0,  \label{dvas}
\end{equation}%
\begin{equation}
a_{-1,t}+a_{0,x}-a_{-2}V_{x}=0,  \label{tris}
\end{equation}%
\begin{equation}
a_{k,t}+a_{k+1,x}+ka_{k-1}V_{x}=0,\ \ k=0,1,\ldots  \label{finito}
\end{equation}%
Integration of (\ref{ras}) yields $a_{-2}=a_{-2}(t)$. However,
without loss of generality one can set $a_{-2}=1$. For this we can
perform the following point transformation of the independent
variables: $(t,x,p)\mapsto (y, z, q)$ with $t=t(y)$,
$x=x(y,z)=a(y)z$, $p=p(y,z,q)=\frac{a^{\prime}(y)}{t^{\prime
}(y)}z+\frac{a(y)}{t^{\prime }(y)}q$. The function $t(y)$ is to be
found from the equation $t'(y) = \big(a_{-2}(t)\big)^2$ and $a(y)=
a_{-2}(t(y))$. Indeed, under this transformation Vlasov kinetic
equation (\ref{vlas-bis})  is transformed into
$F_{y}+qF_{z}-F_{q}W_{z}=0$ with $W=W(y,z)$ found from
$V_{x}+(t'(y))^{-2}x_{yy}-x_{y}(t'(y))^{-3}t''(y)=x_{z}(t'(y))^{-2}W_{z}$
while asymptotic series (\ref{laurent}) becomes
\begin{equation*}
F(y,z,q)=q+\tilde{a}_{-1}(y,z)+\frac{\tilde{a%
}_{0}(y,z)}{q}+\frac{\tilde{a}_{1}(y,z)}{q^{2}}+\ldots
\end{equation*}%
Thus, if we choose $a_{-2}=1$, then (\ref{dvas}) yields $\tilde
a_{-1}=\tilde a_{-1}(y)$. However, we can shift $q$ by the value
$\tilde a_{-1}(y)$. This requires the following transformation for
(\ref{vlas-bis}): $(y, z, q)\mapsto (t=y, x=z + s(y), p=q + \tilde
a_{-1}(y))$, $W(y,z) \mapsto \bar V(x,t)=W(z-s(t),t) +
\tilde{a}'_{-1}(t)\cdot z$ with $s'(t)=a_{-1}(t)$. Then asymptotic
series (\ref{laurent}) assumes the form
\begin{equation*}
F(x,t,p)=p+\frac{\bar a_{0}(x,t)}{p}+\frac{\bar
a_{1}(x,t)}{p^{2}}+\ldots
\end{equation*}%
and (\ref{tris}) reduces to $\bar a_{0,x}=\bar V_{x}$. Since the
potential function $V(x,t)$ is involved in Vlasov kinetic
equation (\ref{vlas-bis}) via its
derivative $V_{x}$, without loss of generality we can choose%
\begin{equation*}
a_{0}=V.
\end{equation*}%
Thus, corresponding infinite set of equations (\ref{finito}) together with
this condition $V=a_{0}$ implies Benney hydrodynamic chain (\ref{cepochka}).

Now we study the pair (\ref{vlas-bis})+(\ref{moment-bis}). We will
prove here that substitution of (\ref{moment-bis}) into
(\ref{cepochka}) implies Vlasov kinetic equation (\ref{vlas-bis})
again. Indeed, at the first step we obtain
\begin{equation*}
\overset{\infty }{\underset{-\infty }{\int }}p^{k}\Phi ^{\prime
}(F)(F_{t}dp+pF_{x})dp+kA_{x}^{0}\overset{\infty }{\underset{-\infty }{\int }%
}p^{k-1}\Phi (F)dp=0.
\end{equation*}%
Integrating by parts we get%
\begin{equation*}
\overset{\infty }{\underset{-\infty }{\int }}p^{k}\Phi ^{\prime
}(F)(F_{t}+pF_{x}-F_{p}A_{x}^{0})dp=0.
\end{equation*}%
Since $k$ is arbitrary, infinite set of these integrals vanish if the
function $F(x,t,p)$ satisfies the Vlasov kinetic equation%
\begin{equation*}
F_{t}+pF_{x}-F_{p}A_{x}^{0}=0,
\end{equation*}%
where according to this procedure%
\begin{equation*}
A^{0}=\overset{\infty }{\underset{-\infty }{\int }}\Phi (F)dp.
\end{equation*}%

As a result of our considerations
we conclude that in fact
the Benney chain (\ref{cepochka}) is the pivotal object relating \emph{%
different} pairs (\ref{vlas-bis})+(\ref{expand-bis}) and (\ref{vlas-bis})+(%
\ref{moment-bis}). Certainly we can make a way through Benney chain from one
pair to another pair. In this way one obtains an interesting transformation.
Namely substitution of (\ref{moment-bis}) into (\ref{expand-bis}) yields a (%
\emph{formal}) integral transformation%
\begin{equation*}
F(x,t,p)=p+\overset{\infty }{\underset{m=0}{\sum }}\frac{A^{m}}{p^{m+1}}=p+%
\overset{\infty }{\underset{m=0}{\sum }}\frac{1}{p^{m+1}}\overset{\infty }{%
\underset{-\infty }{\int }}q^{m}\Phi (\tilde{F}(x,t,q))dq=p+\overset{\infty }%
{\underset{-\infty }{\int }}\frac{\Phi (\tilde{F}(x,t,q))}{p-q}dq,
\end{equation*}%
where $\tilde{F}(x,t,p)$ is a given solution of Vlasov kinetic equation (\ref%
{vlas-bis}), and $F(x,t,p)$ is a new solution. This transformation was
obtained for the Vlasov equation in \cite{GT} and used in hydrodynamics in
\cite{Chesn}.
One can show directly that this transformation maps a solution
$F(x,t,p)$ of Vlasov kinetic equation into another solution of the
same equation if $V(x,t)=\int\Phi (F(x,t,q))dq$. Indeed, suppose
that some $F(x,t,q)$ satisfies Vlasov kinetic equation
(\ref{vlas-bis})
\begin{equation}
F_{t}+qF_{x}-F_{q}V_{x}=0.  \label{odno}
\end{equation}%
Then obviously any function $\Phi (F(x,t,q))$ also satisfies the
same equation. Let us multiply this equation by $(p-q)^{-1}$ and
integrate with respect to $q$ along an arbitrary path $D$:
\begin{equation*}
\underset{D}{\oint }\frac{\big(\Phi (F)\big)_{t}+q\big(\Phi (F)\big)_{x}
-\big(\Phi (F)\big)_{q}V_{x}}{p-q}%
dq=0
\end{equation*}%
or
\begin{equation}
\left( \underset{D}{\oint }\frac{\Phi (F)dq}{p-q}\right) _{t}+\left(
\underset{D}{\oint }\frac{q\Phi (F)dq}{p-q}\right) _{x}-V_{x}\underset{D}{%
\oint }\frac{\Phi _{q}(F)dq}{p-q}=0.  \label{chasti}
\end{equation}%
The second integral can be transformed:
\begin{equation*}
\underset{D}{\oint }\frac{q\Phi (F)dq}{p-q}=p\underset{D}{\oint }\frac{%
\Phi (F)dq}{p-q}-\underset{D}{\oint }\Phi (F)dq,
\end{equation*}%
while the third integral reduces to
\begin{equation*}
\underset{D}{\oint }\frac{\Phi _{q}(F)dq}{p-q}=\left.\frac{\Phi
(F)}{p-q}\right|_{D}- \underset{D}{\oint }\frac{\Phi
(F)dq}{(p-q)^{2}}=\left.\frac{\Phi (F)}{p-q}\right|_{\partial
D}+\left( \underset{D}{\oint }\frac{\Phi (F)dq}{p-q}\right) _{p}.
\end{equation*}%
Introducing new function
\begin{equation}
\tilde{F}(x,t,p)=p+\underset{D}{\oint }\frac{\Phi (F)dq}{p-q},
\label{integral}
\end{equation}%
we can see that (\ref{chasti}) becomes
\begin{equation}
\tilde{F}_{t}+p\tilde{F}_{x}-\left( \underset{D}{\oint }\Phi
(F)dq\right) _{x}=V_{x}\left.\frac{\Phi (F)}{p-q}\right|_{\partial
D}+\tilde{F}_{p}V_{x}-V_{x}.  \label{vmeste}
\end{equation}%
If $\Phi (F)$ vanishes on $\partial D$ (or in the particular case
when the integration is performed along the real axis from $-\infty$
up to $+\infty $, then $\Phi (F)$ must be rapidly decreasing
function at the infinities), and if we set
\begin{equation*}
V(x,t)=\underset{D}{\oint }\Phi (F(x,t,q))dq,
\end{equation*}%
then (\ref{vmeste}) is nothing but the same Vlasov kinetic equation
(cf. (\ref{odno})) with \emph{the same $V(x,t)$}:
\begin{equation*}
\tilde{F}_{t}+p\tilde{F}_{x}-\tilde{F}_{p}V_{x}=0.
\end{equation*}%
Expanding (\ref{integral}) for $p \rightarrow \infty$ we get the
asymptotic expansion (\ref{expand-bis}) where all moments are
determined precisely by (\ref{moment-bis}).

\textbf{Remark}. In some physical applications (for instance in
hydrodynamics, see \cite{Khe}) Vlasov type kinetic equation (\ref{vlas-bis})
derived from some fundamental physical laws contains the potential function $%
V(x,t)$, which is different from $A^{0}$. For instance, $V(x,t)=\ln
A^{0}=\ln \int Fdp$. The corresponding Benney-like hydrodynamic
chain for $A^{k}(x,t)=\int p^{k}\Phi (F(x,t,p))dp$ is
\begin{equation*}
A_{t}^{k}+A_{x}^{k+1}+kA^{k-1}(\ln A^{0})_{x}=0,\text{ \ }k=0,1, \ldots
\end{equation*}%
It is non-integrable by the method of hydrodynamic reductions (see
\cite{GT}). We consider the opposite case in this paper: the
integrable (by the method of hydrodynamic reductions) version of
Vlasov kinetic equation determined by the restriction $V=A^{0}$.

\section*{Appendix B (Egorov Pairs of Conservation Laws)}

\addcontentsline{toc}{section}{Appendix B (Egorov Pairs of
Conservation Laws)}

We will use the techniques of \cite{tsar91,MaksTsar} in order to prove the
result we need in Section~\ref{sec-HRedI} for construction of the basic
formulae for solutions (\ref{hodo}), (\ref{alga}), namely the statement that
\emph{for arbitrarily chosen conservation law density $h(\mathbf{r})$ of the
original Egorov system (in our case (\ref{rim})) an appropriately chosen
commuting flow must have a Egorov pair such that $f_{\tau }=h_{x}$, where $%
f=A^{0}$ (the density $f$ in Lemma~\ref{lemma-Egor}).}

Egorov semi-Hamiltonian hydrodynamic type systems have the following form
\begin{equation}  \label{App-B-sys}
r_{t}^{i}=\frac{\tilde{H}_{i}}{\bar{H}_{i}}r_{x}^{i},
\end{equation}%
where the (non-flat in general) metric is given by $g_{ii}=\bar{H}_{i}^{2}$
and the rotation coefficients are%
\begin{equation*}
\beta _{ik}=\frac{\partial _{i}\bar{H}_{k}}{\bar{H}_{i}}, \quad i \neq k.
\end{equation*}%
The Egorov property for semi-Hamiltonian systems consists in symmetricity of
the rotation coefficients: $\beta _{ik} = \beta _{ki}$. Corresponding linear
system reads%
\begin{equation}  \label{App-B-bijH}
\partial _{i}H_{k}=\beta _{ik}H_{i}, \quad i \neq k.
\end{equation}%
One particular solution of this system is $\tilde{H}_{i}$, another
particular solution is $\bar{H}_{i}$.

We know that Egorov pair $f_{t}=h_{x}$, $h_{t}=g_{x}$ of conservation laws
for (\ref{App-B-sys}) are given by the formulae (\cite{MaksTsar}, Theorem~1)
\begin{equation*}
\partial _{i}f=\bar{H}_{i}^{2},\text{ \ }\partial _{i}h=\bar{H}_{i}\tilde{H}%
_{i}=\tilde{H}_{i}\bar{H}_{i},\text{ \ }\partial _{i}g=\tilde{H}_{i}^{2}.
\end{equation*}%
The conservation law densities for (\ref{App-B-sys}) are given by
\begin{equation}
\partial _{i}\hat{h}=\bar{H}_{i}\hat{H}_{i},  \label{App-B-cons}
\end{equation}%
where $\hat{H}_{i}$ are arbitrary solutions of (\ref{App-B-bijH}). Also, we
know that all commuting flows have the form%
\begin{equation}
r_{\tau }^{i}=\frac{\breve{H}_{i}}{\bar{H}_{i}}r_{x}^{i}  \label{App-B-comm}
\end{equation}%
where $\breve{H}_{i}$ are again arbitrary solutions of (\ref{App-B-bijH}).
The Egorov pair for this commuting flow is given by
\begin{equation*}
\partial _{i}f=\bar{H}_{i}^{2},\text{ \ }\partial _{i}\breve{h}=\bar{H}_{i}%
\breve{H}_{i},\text{ \ }\partial _{i}\breve{g}=\breve{H}_{i}^{2}.
\end{equation*}%
Comparing (\ref{App-B-cons}) and (\ref{App-B-comm}) we see that we can
always choose the same solution $\breve{H}_{i}=\hat{H}_{i}$ of (\ref%
{App-B-bijH}) and obtain the necessary commuting flow (\ref{App-B-comm})
with the required $h$ in its Egorov pair.

\section*{Appendix C (L\"owner Equations)}

\addcontentsline{toc}{section}{Appendix C (L\"owner Equations)}

We start from the Vlasov kinetic equation%
\begin{equation}
F_{t}+pF_{x}-F_{p}V_{x}=0  \label{AppC-1}
\end{equation}%
and hydrodynamic type system%
\begin{equation}
a_{t}^{k}+\left( \frac{(a^{k})^{2}}{2}+V(\mathbf{a})\right) _{x}=0,
\label{AppC-2}
\end{equation}%
such that $F(x,t,p)=\lambda (\mathbf{a}(x,t),p)$ with some fixed $\lambda (%
\mathbf{a},p)$ satisfies (\ref{AppC-1}) for arbitrary solution $a^{i}(x,t)$
of (\ref{AppC-2}). Then we obtain%
\begin{equation*}
\sum_{i}\big(\lambda _{i}a_{t}^{i}+p\lambda _{i}a_{x}^{i}-\lambda
_{p}V_{i}a_{x}^{i}\big)=0.
\end{equation*}%
Here and everywhere below we use the lower indices to denote partial
derivatives w.r.t. $a^{i}$: $\lambda _{i}\equiv \partial _{i}\lambda \equiv
\partial \lambda \big/\partial a^{i}$. Substituting $a_{t}^{i}$ from (\ref%
{AppC-2}) we get
\begin{equation*}
\sum_{i}\lambda _{i}[-a^{i}a_{x}^{i}-V_{x}]+p\sum_{i}\lambda
_{i}a_{x}^{i}-\lambda _{p}V_{x}=0,
\end{equation*}%
or%
\begin{equation*}
p\sum_{i}\lambda _{i}a_{x}^{i}-\sum_{i}\lambda
_{i}a^{i}a_{x}^{i}=\sum_{i}\left( \lambda _{p}+\sum_{m}\lambda _{m}\right)
V_{i}a_{x}^{i}.
\end{equation*}%
Since $a^{i}$ are arbitrary solutions of (\ref{AppC-2}) we conclude that
\begin{equation}
\lambda _{i}=\frac{V_{i}}{p-a^{i}}\left( \lambda _{p}+\sum_{m}\lambda
_{m}\right) .  \label{AppC-3}
\end{equation}%
Summing up we obtain%
\begin{equation*}
\sum_{m}\lambda _{m}=\sum_{n}\frac{V_{n}}{p-a^{n}}\left( \lambda
_{p}+\sum_{m}\lambda _{m}\right)
\end{equation*}%
or%
\begin{equation*}
\sum_{m}\lambda _{m}=\sum_{n}\frac{V_{n}}{p-a^{n}}\lambda _{p}\left(
1-\sum_{n}\frac{V_{n}}{p-a^{n}}\right) ^{-1}.
\end{equation*}%
Substituting this into (\ref{AppC-3}) we get%
\begin{equation*}
\lambda _{i}=\frac{V_{i}}{p-a^{i}}\left( 1+\sum_{n}\frac{V_{n}}{p-a^{n}}%
\left( 1-\sum_{n}\frac{V_{n}}{p-a^{n}}\right) ^{-1}\right) \lambda _{p}
\end{equation*}%
i.e. the required formula
\begin{equation*}
\lambda _{i}=\frac{V_{i}}{p-a^{i}}\left( 1-\sum_{n}\frac{V_{n}}{p-a^{n}}%
\right) ^{-1}\lambda _{p}.
\end{equation*}

\section*{Appendix D (Principal Series of Conservation Laws)}

\addcontentsline{toc}{section}{Appendix D (Principal Series of
Conservation Laws)}

In this Appendix we prove that one can find all principal series $p_{k}^{i}(%
\mathbf{a})$ of conservation law densities in the expansion
\begin{equation}
p^{(i)}(\mathbf{a},\tilde{\lambda}_{(k)})=a^{i}+p_{1}^{i}(\mathbf{a})\tilde{%
\lambda}_{(k)}+p_{2}^{i}(\mathbf{a})\tilde{\lambda}_{(k)}^{2}+p_{3}^{i}(%
\mathbf{a})\tilde{\lambda}_{(k)}^{3}+\ldots ,\text{ \ }i=1,\ldots ,N,
\label{AppD-princ}
\end{equation}%
where (unknown at this point) generating function $p(\mathbf{a},\lambda )$
of conservation law densities should satisfy
\begin{equation}
p_{t}+\left( \frac{p^{2}}{2}+V(\mathbf{a})\right) _{x}=0  \label{AppD-genred}
\end{equation}%
and the potential function $V(\mathbf{a})$ is already found (as a solution
of Gibbons-Tsarev equations (\ref{gibtsar})). Substitution of (\ref%
{AppD-princ}) into (\ref{AppD-genred}) yields infinite set of equations%
\begin{equation}
a_{t}^{i}+\left( \frac{(a^{i})^{2}}{2}+V(\mathbf{a})\right) _{x}=0,
\label{AppD-a}
\end{equation}%
\begin{equation}
(p_{1}^{i}(\mathbf{a}))_{t}+(a^{i}p_{1}^{i}(\mathbf{a}))_{x}=0,
\label{AppD-b}
\end{equation}%
\begin{equation}
(p_{2}^{i}(\mathbf{a}))_{t}+\left( a^{i}p_{2}^{i}(\mathbf{a})+\frac{1}{2}%
(p_{1}^{i}(\mathbf{a}))^{2}\right) _{x}=0,\ldots .  \label{AppD-g}
\end{equation}%
We will show that all conservation law densities $p_{m}^{i}(\mathbf{a})$ can
be found in quadratures in the first case $p_{1}^{i}(\mathbf{a})$. The
higher elements of the principal series $p_{m}^{i}(\mathbf{a})$ are found in
the same way. First, we observe that (\ref{AppD-a}) coincides with (\ref%
{semi}). Equations (\ref{AppD-b}) give
\begin{equation*}
\sum_{k}\partial _{k}p_{1}^{i}(\mathbf{a})a_{t}^{k}+p_{1}^{i}(\mathbf{a}%
)a_{x}^{i}+\sum_{k}a^{i}\partial _{k}p_{1}^{i}(\mathbf{a})a_{x}^{k}=0.
\end{equation*}%
Substitution of $a_{t}^{k}$ from (\ref{AppD-a}) gives
\begin{equation*}
\sum_{k}\partial _{k}p_{1}^{i}(\mathbf{a})[a^{k}a_{x}^{k}+V_{x}]=p_{1}^{i}(%
\mathbf{a})a_{x}^{i}+\sum_{k}a^{i}\partial _{k}p_{1}^{i}(\mathbf{a}%
)a_{x}^{k},
\end{equation*}%
or
\begin{equation*}
\sum_{k}\big(a^{k}\partial _{k}p_{1}^{i}(\mathbf{a})a_{x}^{k}+(\delta
p_{1}^{i}(\mathbf{a}))\partial _{k}Va_{x}^{k}\big)=p_{1}^{i}(\mathbf{a}%
)a_{x}^{i}+\sum_{k}a^{i}\partial _{k}p_{1}^{i}(\mathbf{a})a_{x}^{k},
\end{equation*}%
where $\delta =\sum_{m}\partial /\partial a^{m}$. Since $a^{s}(x,t)$ are
arbitrary solutions of (\ref{AppD-a}), coefficients at $a_{x}^{s}$ vanish
identically. For $s=i$ this gives us
\begin{equation}
a^{i}\partial _{i}p_{1}^{i}(\mathbf{a})+\delta p_{1}^{i}(\mathbf{a})\cdot
\partial _{i}V=p_{1}^{i}(\mathbf{a})+a^{i}\partial _{i}p_{1}^{i}(\mathbf{a}).
\label{AppD-c}
\end{equation}%
If $s\neq i$ then
\begin{equation}
a^{k}\partial _{k}p_{1}^{i}(\mathbf{a})+\delta p_{1}^{i}(\mathbf{a})\cdot
\partial _{k}V=a^{i}\partial _{k}p_{1}^{i}(\mathbf{a}).  \label{AppD-d}
\end{equation}%
Equation (\ref{AppD-d}) simplifies to the form%
\begin{equation}
\partial _{k}p_{1}^{i}(\mathbf{a})=\delta p_{1}^{i}(\mathbf{a})\frac{%
\partial _{k}V}{a^{i}-a^{k}},\quad k\neq i,  \label{AppD-e}
\end{equation}%
while (\ref{AppD-c}) is%
\begin{equation}
\delta \ln p_{1}^{i}(\mathbf{a})=\frac{1}{\partial _{i}V}.  \label{AppD-f}
\end{equation}%
Equation (\ref{AppD-e}) after summation has the form%
\begin{equation*}
\delta p_{1}^{i}(\mathbf{a})=\delta p_{1}^{i}(\mathbf{a})\underset{m\neq i}{%
\sum }\frac{\partial _{m}V}{a^{i}-a^{m}}+\partial _{i}p_{1}^{i}(\mathbf{a})
\end{equation*}%
or
\begin{equation*}
\delta \ln p_{1}^{i}(\mathbf{a})\left( 1-\underset{m\neq i}{\sum }\frac{%
\partial _{m}V}{a^{i}-a^{m}}\right) =\partial _{i}\ln p_{1}^{i}(\mathbf{a}).
\end{equation*}%
Then taking into account (\ref{AppD-f}), we obtain%
\begin{equation}
\partial _{i}\ln p_{1}^{i}(\mathbf{a})=\frac{1}{\partial _{i}V}\left( 1-%
\underset{m\neq i}{\sum }\frac{\partial _{m}V}{a^{i}-a^{m}}\right) .
\label{AppD-h}
\end{equation}%
Then (\ref{AppD-e}) takes the form%
\begin{equation*}
\partial _{k}\ln p_{1}^{i}(\mathbf{a})=\delta \ln p_{1}^{i}(\mathbf{a})\frac{%
\partial _{k}V}{a^{i}-a^{k}},\quad k\neq i,
\end{equation*}%
and taking into account (\ref{AppD-f}), we obtain%
\begin{equation*}
\partial _{k}\ln p_{1}^{i}(\mathbf{a})=\frac{1}{\partial _{i}V}\frac{%
\partial _{k}V}{a^{i}-a^{k}},\quad k\neq i.
\end{equation*}%
So the conservation law densities $p_{1}^{i}(\mathbf{a})$ can be found in
quadratures:%
\begin{equation*}
d\ln p_{1}^{i}(\mathbf{a})=\frac{1}{\partial _{i}V}\left( 1-\underset{m\neq i%
}{\sum }\frac{\partial _{m}V}{a^{i}-a^{m}}\right) da^{i}+\frac{1}{\partial
_{i}V}\underset{m\neq i}{\sum }\frac{\partial _{m}V}{a^{i}-a^{m}}da^{m}.
\end{equation*}

\textbf{Example:} if $V=\sum_m \epsilon _{m}a^{m}$, then%
\begin{equation*}
d\ln p_{1}^{i}(\mathbf{a})=\frac{1}{\epsilon _{i}}\left( 1-\underset{m\neq i}%
{\sum }\frac{\epsilon _{m}}{a^{i}-a^{m}}\right) da^{i}+\frac{1}{\epsilon _{i}%
}\underset{m\neq i}{\sum }\frac{\epsilon _{m}}{a^{i}-a^{m}}da^{m},
\end{equation*}%
so
\begin{equation*}
p_{1}^{i}(\mathbf{a})=e^{\frac{a^{i}}{\epsilon _{i}}}\underset{m\neq i}{%
\prod }(a^{i}-a^{m})^{-\frac{\epsilon _{m}}{\epsilon _{i}}}.
\end{equation*}

All higher conservation law densities can be found in the same way. For
instance, (\ref{AppD-g}) leads to%
\begin{equation*}
\sum_{k}\partial _{k}p_{2}^{i}(\mathbf{a})a_{t}^{k}+p_{2}^{i}(\mathbf{a}%
)a_{x}^{i}+\sum_{k}\left( a^{i}\partial _{k}p_{2}^{i}(\mathbf{a}%
)a_{x}^{k}+p_{1}^{i}(\mathbf{a})\partial _{k}p_{1}^{i}(\mathbf{a}%
)a_{x}^{k}\right) =0.
\end{equation*}%
Taking into account (\ref{AppD-a}) again, we obtain%
\begin{equation*}
\sum_{k}\partial _{k}p_{2}^{i}(\mathbf{a})[a^{k}a_{x}^{k}+V_{x}]=p_{2}^{i}(%
\mathbf{a})a_{x}^{i}+\sum_{k}\left( a^{i}\partial _{k}p_{2}^{i}(\mathbf{a}%
)a_{x}^{k}+p_{1}^{i}(\mathbf{a})\partial _{k}p_{1}^{i}(\mathbf{a}%
)a_{x}^{k}\right) .
\end{equation*}%
Then%
\begin{equation*}
\sum_{k}\left( a^{k}\partial _{k}p_{2}^{i}(\mathbf{a})a_{x}^{k}+\delta
p_{2}^{i}(\mathbf{a})\partial _{k}Va_{x}^{k}\right) =p_{2}^{i}(\mathbf{a}%
)a_{x}^{i}+\sum_{k}\left( a^{i}\partial _{k}p_{2}^{i}(\mathbf{a}%
)a_{x}^{k}+p_{1}^{i}(\mathbf{a})\partial _{k}p_{1}^{i}(\mathbf{a}%
)a_{x}^{k}\right) .
\end{equation*}%
If $k\neq i$
\begin{equation}
\partial _{k}p_{2}^{i}(\mathbf{a})=\delta p_{2}^{i}(\mathbf{a})\frac{%
\partial _{k}V}{a^{i}-a^{k}}-\frac{p_{1}^{i}(\mathbf{a})\partial
_{k}p_{1}^{i}(\mathbf{a})}{a^{i}-a^{k}}.  \label{level2}
\end{equation}%
If $k=i$
\begin{equation*}
\delta p_{2}^{i}(\mathbf{a})\partial _{i}V=p_{2}^{i}(\mathbf{a})+p_{1}^{i}(%
\mathbf{a})\partial _{i}p_{1}^{i}(\mathbf{a}).
\end{equation*}%
Then%
\begin{equation}
\delta p_{2}^{i}(\mathbf{a})=\frac{1}{\partial _{i}V}p_{2}^{i}(\mathbf{a})+%
\frac{1}{\partial _{i}V}p_{1}^{i}(\mathbf{a})\partial _{i}p_{1}^{i}(\mathbf{a%
})  \label{lev2}
\end{equation}%
and (\ref{level2}) assumes the form%
\begin{equation}
\partial _{k}p_{2}^{i}(\mathbf{a})=\frac{1}{\partial _{i}V}\frac{\partial
_{k}V}{a^{i}-a^{k}}p_{2}^{i}(\mathbf{a})+\frac{1}{\partial _{i}V}\frac{%
\partial _{k}V}{a^{i}-a^{k}}p_{1}^{i}(\mathbf{a})\partial _{i}p_{1}^{i}(%
\mathbf{a})-\frac{p_{1}^{i}(\mathbf{a})\partial _{k}p_{1}^{i}(\mathbf{a})}{%
a^{i}-a^{k}}.  \label{lev3}
\end{equation}%
Let us introduce intermediate set of functions $q_{2}^{i}(\mathbf{a})$ such
that $p_{2}^{i}(\mathbf{a})=q_{2}^{i}(\mathbf{a})p_{1}^{i}(\mathbf{a})$.
Then (\ref{lev2}) by virtue of (\ref{AppD-f}) reduces to the form%
\begin{equation}
\delta q_{2}^{i}(\mathbf{a})=\frac{1}{\partial _{i}V}\partial _{i}p_{1}^{i}(%
\mathbf{a}),  \label{AppD-j}
\end{equation}%
while (\ref{lev3}) due to (\ref{AppD-h}) implies:%
\begin{equation*}
\partial _{k}q_{2}^{i}(\mathbf{a})=\frac{1}{\partial _{i}V}\frac{\partial
_{k}V}{a^{i}-a^{k}}\partial _{i}p_{1}^{i}(\mathbf{a})-\frac{\partial
_{k}p_{1}^{i}(\mathbf{a})}{a^{i}-a^{k}},\quad k\neq i.
\end{equation*}%
This equation after summation has the form%
\begin{equation*}
\delta q_{2}^{i}(\mathbf{a})=\underset{m\neq i}{\sum }\frac{1}{\partial _{i}V%
}\frac{\partial _{m}V}{a^{i}-a^{m}}\partial _{i}p_{1}^{i}(\mathbf{a})-%
\underset{m\neq i}{\sum }\frac{\partial _{m}p_{1}^{i}(\mathbf{a})}{%
a^{i}-a^{m}}+\partial _{i}q_{2}^{i}(\mathbf{a}).
\end{equation*}%
Then taking into account (\ref{AppD-j}), we obtain%
\begin{equation*}
\partial _{i}q_{2}^{i}(\mathbf{a})=\frac{1}{\partial _{i}V}\partial
_{i}p_{1}^{i}(\mathbf{a})-\underset{m\neq i}{\sum }\frac{1}{\partial _{i}V}%
\frac{\partial _{m}V}{a^{i}-a^{m}}\partial _{i}p_{1}^{i}(\mathbf{a})+%
\underset{m\neq i}{\sum }\frac{\partial _{m}p_{1}^{i}(\mathbf{a})}{%
a^{i}-a^{m}}.
\end{equation*}%
Thus, $q_{2}^{i}(\mathbf{a})$ can be found in quadratures:%
\begin{eqnarray}
dq_{2}^{i}(\mathbf{a}) &=&\left( \frac{1}{\partial _{i}V}\partial
_{i}p_{1}^{i}(\mathbf{a})-\underset{m\neq i}{\sum }\frac{1}{\partial _{i}V}%
\frac{\partial _{m}V}{a^{i}-a^{m}}\partial _{i}p_{1}^{i}(\mathbf{a})+%
\underset{m\neq i}{\sum }\frac{\partial _{m}p_{1}^{i}(\mathbf{a})}{%
a^{i}-a^{m}}\right) da^{i}+{}  \notag \\
&&{}+\underset{m\neq i}{\sum }\left( \frac{1}{\partial _{i}V}\frac{\partial
_{m}V}{a^{i}-a^{m}}\partial _{i}p_{1}^{i}(\mathbf{a})-\frac{\partial
_{m}p_{1}^{i}(\mathbf{a})}{a^{i}-a^{m}}\right) da^{m}.  \notag
\end{eqnarray}

\section*{Acknowledgements}

Our special thanks to our Scientific Advisor Professor Sergey
Petrovich Novikov for the immense contribution of his groundbreaking
ideas to our results and his permanent support and attention.
Authors thank S.P. Novikov, B.A. Dubrovin, E.V. Ferapontov, P.G.
Grinevich, B.G.~Konopelchenko, O.I.~Morozov and V.V. Vedenyapin for
their stimulating and clarifying discussions.

MVP's work was partially supported by the RF Government grant
\#2010-220-01-077, ag. \#11.G34.31.0005, by the grant of Presidium of RAS
\textquotedblleft Fundamental Problems of Nonlinear
Dynamics\textquotedblright\ and by the RFBR grant 11-01-00197.

\end{document}